\documentclass[a4paper, oneside, fleqn, 12pt]{scrartcl}

\usepackage[ngerman,USenglish]{babel}

\usepackage[authoryear,round,colon]{natbib}

\providecommand{\textcite}{\citet}

\usepackage[leqno]{amsmath}
\usepackage{amssymb}
\usepackage{textcomp}
\usepackage{dsfont}

\usepackage{libertinus} 
\usepackage{bm}        
\usepackage{microtype}

\usepackage{graphicx}
\usepackage{multirow}
\usepackage{pdflscape}
\usepackage{tabularx}
\usepackage{tabulary}
\usepackage{booktabs}
\usepackage[font={small,sl},labelfont=bf]{caption}

\usepackage{setspace}
\usepackage{array}
\usepackage[title]{appendix}
\usepackage[running]{lineno}
\usepackage[marginparwidth=2.5cm]{geometry}
\usepackage{csquotes}
\usepackage{siunitx}
\DeclareSIUnit\knot{kn}
\usepackage{xcolor}

\usepackage{authblk}

\usepackage{hyperref}
\hypersetup{colorlinks=true,
    linkcolor=blue,
    citecolor=blue,
    bookmarksopen=true,
    pdfstartpage={1},
    pdftitle={Evaluating LiDAR Data Sources, Predictor Resolution, and Spatial Random Effects in Bayesian Change-of-Support Models for Forest Inventory},
    pdfauthor={Arthur Spitta, Andrew O. Finley, Christoph Gollob, Ralf Kraßnitzer, Tim Ritter, Andreas Tockner,Arne Nothdurft},
    pdfkeywords={LiDAR, forest inventory, model-based small area estimation, UAV, spatial Bayesian regression models}
}

\graphicspath{{figures/}}

\newcommand{\bsy}{\bm{y}}
\newcommand{\bsx}{\bm{x}}
\newcommand{\bsb}{\bm{\beta}}
\newcommand{\bsw}{\bm{\omega}}
\newcommand{\bse}{\bm{\epsilon}}
\newcommand{\bstheta}{\bm{\theta}}

\title{Evaluating LiDAR Data Sources, Predictor Resolution, and Spatial Random Effects in Bayesian Change-of-Support Models for Forest Inventory}

\author[a]{\large Arthur Spitta\thanks{\href{mailto:arthur@spitta.org}{arthur@spitta.org}, Tel: +49 1575 2597815}}
\author[b]{\large Andrew O. Finley}
\author[a]{\large Christoph Gollob}
\author[a]{\large Ralf Kraßnitzer}
\author[a]{\large Tim Ritter}
\author[a]{\large Andreas Tockner}
\author[a]{\large Arne Nothdurft}

\affil[a]{\normalsize Department of Ecosystem Management, Climate and Biodiversity, Institute of Forest Growth, BOKU University, Vienna, Austria}
\affil[b]{\normalsize Departments of Forestry and Statistics \& Probability, Michigan State University, USA}

\date{\large \today}

\begin{document}
\maketitle


\begin{abstract}
Forest managers need timely stand-level information to support operational planning, particularly in mixed-species, structurally heterogeneous forests facing climate-related disturbance. Model-based estimation combines sparse field data with remotely sensed predictors to estimate growing stock volume (GSV) for small areas. While uncrewed aerial vehicle laser scanning (ULS) offers flexible, high-resolution LiDAR acquisition, its advantages over conventional airborne laser scanning (ALS) remain unclear.

We compared publicly available ALS and newly acquired ULS data using Bayesian change-of-support models to estimate GSV in a mixed-species forest in north-eastern Germany. We evaluated distributional LiDAR metrics and spatial random effects.

ULS consistently outperformed ALS, achieving cross-validated RMSPEs of \qty{68.5}{\cubic\meter\per\hectare} and \qty{79.5}{\cubic\meter\per\hectare}, respectively—a \qty{13.8}{\percent} reduction in prediction error. Distributional metrics improved ULS models more strongly, reducing RMSPE by up to \qty{10.1}{\percent}; spatial effects provided only minor gains at substantially higher computational cost. ULS also produced lower uncertainty in latent stand-mean GSV estimates.

The ULS advantage may reflect both finer-scale canopy information and closer temporal alignment with field measurements. Timely, information-rich LiDAR may therefore be more valuable for stand-level GSV estimation than increasingly complex spatial models. Temporally matched ALS–ULS comparisons are needed to isolate platform effects.
\end{abstract}

\section{Introduction}
Forest inventories provide the basis for sustainable planning and management in forest operations. In Germany and Austria, operational inventories are commonly conducted at the level of individual management units, or stands, using expert assessments or sample-based methods such as angle-count sampling, and are typically repeated at intervals of 10 years or more \citep[pp.~518, 534--537]{kurth_forsteinrichtung_2018}. Under increasing disturbance pressure and changing forest dynamics associated with climate change, forest inventories may become outdated well before their intended revision, particularly where harvesting and disturbance events substantially alter forest structure, potentially giving rise to the need for more frequent updating \citep{coops_framework_2023, fassnacht_remote_2024, fassnacht_forest_2025}. At the same time, forest management increasingly considers multiple objectives beyond timber production, including carbon storage, biodiversity conservation, and climate-change adaptation, increasing the demand for timely information on forest condition and structure. Growing stock volume (GSV) is a fundamental forest inventory attribute used to characterize forest resources and inform forest management \citep{gschwantner_growing_2022}. It provides information on stand development and changes in woody biomass resulting from growth, mortality, harvesting, and disturbance, forming an important basis for estimating aboveground biomass and forest carbon stocks. 

Population-level sampling designs can provide cost-effective and statistically grounded estimates of forest attributes including GSV for medium to large enterprises and, where sample sizes are sufficient, for larger subpopulations or strata. However, such designs usually do not provide the sampling density required for reliable stand-level estimation under purely design-based inference \citep{dettmann_review_2022}. Although population- or strata-level information may be sufficient for medium- to long-term planning, stand-level information is often preferred by forest managers, particularly for short-term operational planning \citep{ulvdal_handling_2023}. In some cases, it is also a legal requirement, e.g., for German private forest enterprises, for the bottom-up calculation of sustainable harvest rates \citep{bundesministerium_der_finanzen_richtlinien_2017}. 

Remote sensing provides an effective means to bridge the gap between statistically efficient yet sparse sample inventories and the need for spatially explicit stand-level information. In particular, airborne laser scanning (ALS) has become a key source of auxiliary data, as LiDAR-derived canopy height and structural metrics correlate strongly with forest attributes such as GSV \citep{naesset_predicting_2002, wulder_lidar_2012, Nothdurft_2021}. Model-based estimators can seamlessly integrate these auxiliary data into the inventory framework, enabling reliable predictions for areas where sample sizes are too small to support precise design-based or model-assisted estimates \citep{dettmann_review_2022}. Generated at fine spatial resolutions, such predictions can be aggregated into small-area estimates for management units or individual stands \citep{Nothdurft_2025}. Small-area estimation methods differ fundamentally in their inferential framework \citep{Stahl_2016}: while design-based \citep{Mandallaz_2013} and model-assisted approaches \citep{Sarndal_1992} rely strictly on the sampling design for unbiasedness, model-based estimators \citep{Breidenbach_2012, Saarela_2018} depend on the validity of an assumed statistical model. Consequently, model-based inference allows for spatially explicit estimations in sparsely sampled or completely unsampled stands, provided the model is correctly specified and calibrated on representative data.

The practical utility of model-based stand-level estimation depends heavily on the availability, spatial resolution, and temporal consistency of suitable auxiliary data. In this context, UAV-borne laser scanning (ULS) has emerged as a flexible alternative to conventional airborne laser scanning (ALS). By reducing reliance on crewed aircraft operations, ULS enables forest managers to acquire remote sensing data independently and on demand \citep{vandendaele_estimation_2021, ecke_uav-based_2022}. This operational flexibility is particularly crucial in disturbance-prone forests, where canopy structure changes rapidly and close temporal alignment between field measurements and remote sensing acquisitions is essential to minimize temporal mismatch errors. However, these operational benefits are offset by practical constraints, including strict regulatory frameworks, site accessibility issues, limited coverage per flight \citep{puliti_use_2017}, and substantial upfront investments in hardware, certification, and processing pipelines \citep{scudder_cost-effective_2025}.

Beyond these operational trade-offs, ULS offers distinct structural data characteristics. Operating at lower flight altitudes and speeds, ULS provides point densities up to several thousand points per square meter, effectively bridging the resolution gap between regional ALS surveys and localized terrestrial laser scanning (TLS) \citep{bruggisser_comparison_2019}. While recent methodological research has heavily leveraged these ultra-dense point clouds for fine-scale, individual-tree applications, such as annual growth trajectory reconstruction \citep{Puliti_2023} or automated crown segmentation via deep learning \citep{Straker_2023}, their broader utility for area-based forest inventory lies in capturing detailed vertical and horizontal stand structures without requiring heuristic parameterization.

Despite these technological advances, direct comparisons between ULS and ALS as alternative auxiliary data sources within model-based estimation frameworks remain scarce. Most ULS studies to date have focused either on image-based structure-from-motion approaches \citep{dainelli_recent_2021}, individual-tree attribute estimation, or pure platform comparisons based on point cloud metrics \citep{vandendaele_estimation_2021, brede_non-destructive_2019, bruggisser_comparison_2019}. Only a few studies have evaluated ULS within an explicit inferential context; for example, \citet{puliti_comparison_2020} demonstrated the utility of ULS for volume estimation, but their scope was limited to a homogeneous \textit{Pinus radiata} plantation. Similarly, \citet{mcclelland_manned_2019} compared ULS and ALS for carbon stock prediction using functional models, yet technical acquisition issues limited their findings. However, forests managed under increasingly diverse objectives and changing environmental conditions are often characterized by mixed species composition and heterogeneous stand structures. Such structural and compositional heterogeneity may challenge relationships between remotely sensed structural metrics and GSV established in relatively homogeneous stands. Consequently, there is a critical knowledge gap regarding how ULS compares to ALS as an auxiliary data source for model-based inference, particularly in structurally complex forest ecosystems.

\begin{figure}
    \centering
    \includegraphics[width=1\linewidth]{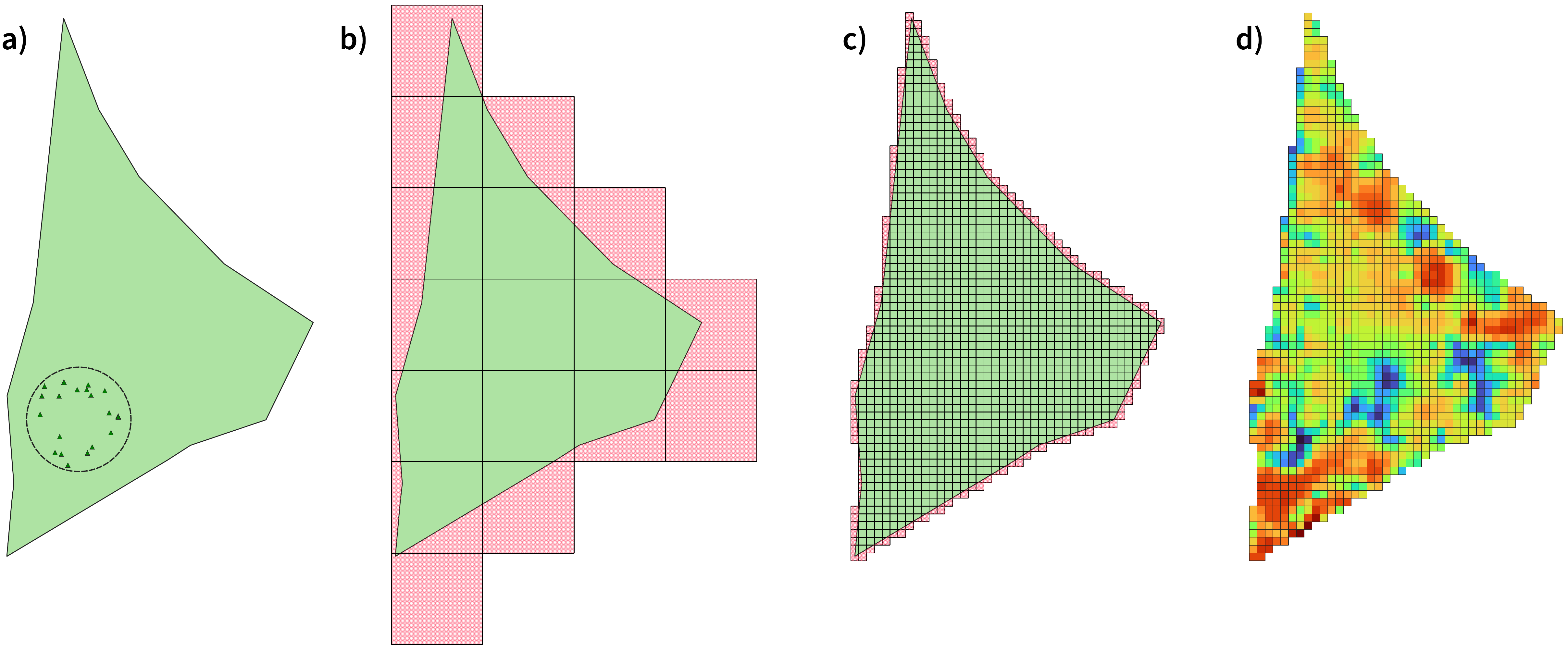}
    \caption{Change of support in model-based forest inventory. \textbf{a)} A forest attribute $y(B_l)$ is observed over a circular field plot $B_l$. \textbf{b)} Raster cells with a support comparable to the field plots result in substantial boundary overlap when predicting for irregular and small target units $U_k$. \textbf{c)} Finer raster cells $A_i$ reduce boundary overlap with stand polygons but have a substantially smaller support than the field plots. \textbf{d)} Remote sensing provides auxiliary information $x(A_i)$ at fine raster-cell support. Because the response is observed at plot support $B_l$, relating $y(B_l)$ to $x(A_i)$ requires accounting for the difference in spatial support when predicting at $A_i$.}
    \label{fig:cos_issue}
\end{figure}

When applying such high-resolution remote sensing data for model-based inference, an inherent spatial challenge arises: while ULS enables predictions at fine spatial scales, field plot sizes remain constrained by operational and measurement considerations. Larger sample plots are generally preferred to minimize plot-level random errors stemming from edge effects, tree location uncertainty, and positioning inaccuracies \citep{white_best_2013}. This mismatch between the spatial support of field observations and prediction units introduces a classic Change-of-Support (COS) problem (Figure~\ref{fig:cos_issue}), defined by differences in the size, shape, and spatial orientation of measurement regions \citep{gotway_combining_2002}. 

In forest inventory applications, three distinct aspects of COS must be addressed. First, support shape differs between circular field plots and square prediction cells; however, when both supports share equal area, shape mismatch has been shown to have negligible practical effects \citep{packalen_circular_2023, strunk_mitigation_2025}. Second, areal overlap occurs because target units (e.g., forest stands) are irregular polygons, whereas prediction rasters are regular grids. Large raster cells overlapping stand boundaries introduce boundary effects and sub-grid heterogeneity. While increasing the spatial resolution of prediction rasters mitigates these boundary errors, it triggers a third COS issue: scale effects. Because statistical relationships established at the field-plot scale do not automatically hold at finer prediction resolutions, standard model predictions cannot simply be applied to arbitrarily small raster cells.

To resolve these spatial scale incompatibilities, \citet{zhang_bayesian_2024} introduced a Bayesian latent-process framework that predicts at a fine spatial resolution before aggregating predictions to target polygons. Building upon this approach, the present study evaluates Bayesian COS regression to optimize stand-level GSV estimation. Beyond resolving spatial support incompatibilities, effective model-based inventory relies on optimizing sensor choice, predictor complexity, and spatial dependency structures. Specifically, this study evaluates:
\begin{enumerate}
    \item the predictive performance of ULS data relative to ALS as alternative auxiliary sources,
    \item the capacity of Bayesian change-of-support models to reduce boundary and scale-related errors through fine-resolution prediction rasters,
    \item the added value of complex structural LiDAR metrics beyond standard canopy height metrics, and
    \item the gain in predictive accuracy achieved by spatial Bayesian modeling over non-spatial regression.
\end{enumerate}
This study aims to provide practical guidance for operational decisions regarding data acquisition and model specification in LiDAR-assisted forest inventories, by jointly evaluating these sensor, spatial, and modeling dimensions.

\section{Materials and Methods}
\subsection{Study region and data collection}

\begin{figure}
    \centering
    \includegraphics[width=1\linewidth]{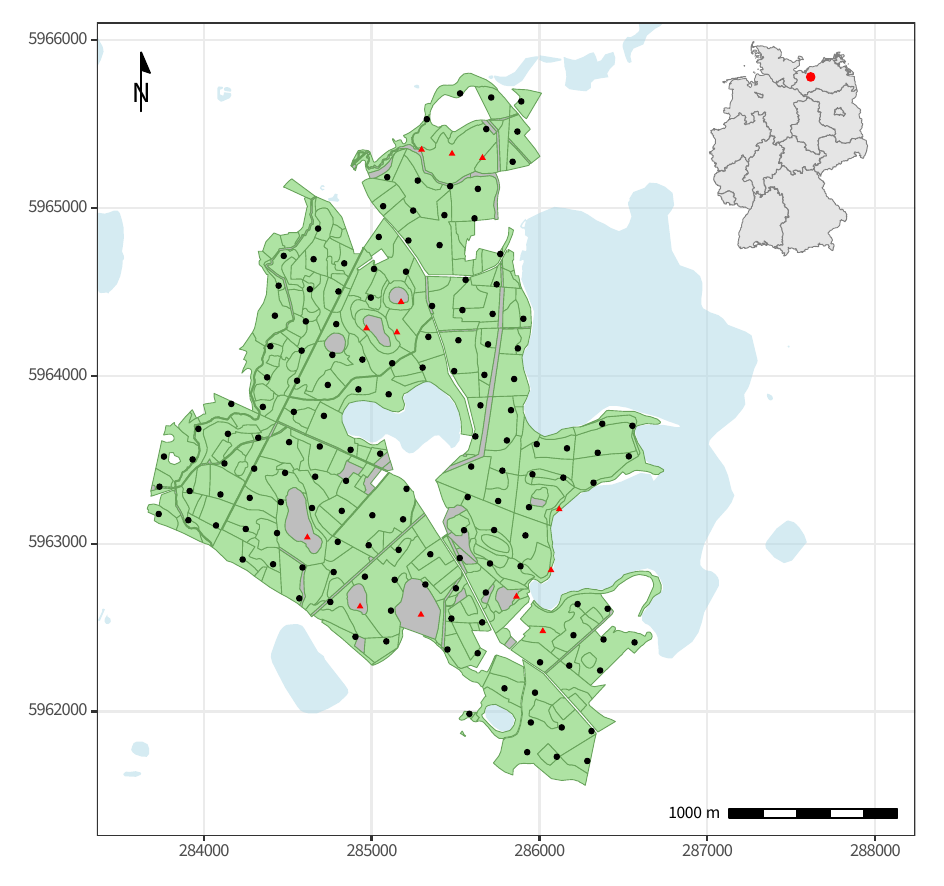}
    \caption{Study location showing management units (green: forest stands, gray: non-forest operational areas) and inventory plot locations (black circles: measured plots, red triangles: skipped plots). Administrative boundaries: geoBoundaries.org, based on data from the Federal Agency for Cartography and Geodesy (BKG), 2021; water bodies: OpenStreetMap contributors, retrieved via Overpass Turbo. Coordinate reference system: ETRS89 / UTM zone 33N}
    \label{fig:map_overview}
\end{figure}

The study region is located in the German federal state of Mecklenburg–West Pomerania, between the villages of Warin and Blankenberg (Figure~\ref{fig:map_overview}). The forest area is divided into \num{311} stands covering a total of \qty{553.5}{\hectare}. Of these, \num{249} stands, covering \qty{513.1}{\hectare}, are classified as forest stands, whereas the remaining stands represent operational areas without tree cover, such as roads, wood storage areas, and administrative buildings. Forest stand sizes range from \qty{0.06}{\hectare} to \qty{17.77}{\hectare}, with a mean of \qty{2.06}{\hectare}, a median of \qty{1.51}{\hectare}, and a standard deviation of \qty{2.00}{\hectare}. The area lies within the young moraine landscape of northern Germany, where glacial deposits form the dominant parent material. The resulting soils are generally well-drained and nutrient-rich, whereas wetter site conditions occur mainly in lower-lying terrain and around lakes. The forest comprises mixed-species stands dominated by European beech (\textit{Fagus sylvatica}) and Scots pine (\textit{Pinus sylvestris}), alongside Douglas fir (\textit{Pseudotsuga menziesii}), pedunculate oak (\textit{Quercus robur}), black alder (\textit{Alnus glutinosa}), Norway spruce (\textit{Picea abies}), and other tree species in varying proportions.

Data collection was based on an initial systematic sampling grid (\qty{183}{\meter} $\times$ \qty{183}{\meter}) with a randomized offset and rotation, yielding a sampling frame of \num{165} potential sample plots across all stands (Figure~\ref{fig:map_overview}). Because \num{14} plots were inaccessible in the field, inventory data were collected on the remaining $n = 151$ plots. Field measurements were conducted from 17 to 23 March 2026 under leaf-off canopy conditions using a handheld mobile personal laser scanner (PLS; GeoSLAM ZEB Horizon, FARO Technologies Inc., Lake Mary, USA) on circular plots with a radius of \qty{20}{\meter}. Plot center points were georeferenced using real-time kinematic global navigation satellite system (RTK-GNSS) measurements, with real-time corrections provided by the regional SAPOS service. Tree species were recorded manually for all measured trees exceeding a diameter at breast height (DBH; measured at \qty{1.30}{\meter} above ground) threshold of \qty{10}{\cm}.

\subsection{Field inventory data}

Tree locations, DBH, and height for \num{6722} trees with DBH $\ge \qty{10}{\cm}$ were automatically extracted from the 3D PLS point clouds using the R package \texttt{treeX} \citep{tockner_automatic_2022}. All individual tree point clouds were visually inspected for segmentation artifacts. For \num{579} trees where segmentation errors compromised the LiDAR-derived height, total tree height was instead predicted from DBH using Prodan's generalized height--diameter function,
\begin{equation}
h_{ij}=\qty{1.3}{\metre}+\frac{d_{ij}^2}{a + b d_{ij} + (c + u_j)d_{ij}^2},\qquad u_j \sim \mathcal{N}(0, \sigma_c^2),
\label{eq:hd_Prodan}
\end{equation}
where $h_{ij}$ is tree height (\unit{\meter}), $d_{ij}$ is diameter at breast height (\unit{\cm}), and $u_j$ is a plot-level random effect on parameter $c$. Models were fitted separately for each tree species group using all uncorrupted tree measurements. Individual stem volume was calculated as merchantable timber volume above a \qty{7}{\centi\meter} top diameter over bark (Vfm) using species-specific taper functions implemented in the R package \texttt{TapeS} \citep{r_tapes}. The \texttt{TapeS} R package extends and adapts the \texttt{TapeR} framework by providing taper-curve models developed from extensive tree measurements collected across several German forest research stations. For willow trees ($n = 33$), where taper functions yielded erratic estimates, stem volume was calculated using a constant form factor of \num{0.3}.

\subsection{LiDAR data acquisition and preprocessing}

UAV-borne laser scanning (ULS) was conducted using a DJI Zenmuse L2 sensor mounted on a DJI Matrice 300 RTK platform (SZ DJI Technology Co., Ltd., Shenzhen, China). In total, 41 flights were acquired between 2 and 4 March 2026 in linear scan mode. The mean flight height was \qty{96}{\meter} above ground level, with a mean ground speed of \qty{9.8}{\meter\per\second}. The sensor was operated with a pulse repetition rate of \qty{240}{\kilo\hertz} and a scan rate of \qty{1200}{lines\per\second}. The nominal field of view was $\ang{70} \times \ang{3}$, with a nominal nadir laser footprint of approximately \qtyproduct{3.8 x 11.5}{\centi\meter}. The mean flight-strip overlap was \qty{16.63}{\meter}. Following trajectory and point-cloud optimization in DJI Terra, the point cloud was randomly subsampled to \qty{25}{\percent} of its original density to reduce computational requirements. This reduction was considered acceptable because the retained point density remained high for derivation of the LiDAR metrics. Within forested areas, the resulting ULS point cloud had a mean density of \qty{284.1}{points\per\square\meter}.

Airborne laser scanning (ALS) data were provided by the state authorities of Mecklenburg–Western Pomerania. The ALS campaign was flown on 16 and 17 March 2023 using a crewed aircraft equipped with a RIEGL VQ-780 II-S sensor (RIEGL Laser Measurement Systems GmbH, Horn, Austria). Flights were conducted at a mean altitude of \qty{1000}{\meter} above ground level and an airspeed of \qty{140}{\knot} ($\approx\qty{72.0}{\meter\per\second}$). The system operated with a pulse repetition rate of \qty{1}{\mega\hertz}, a scan angle of $\pm\ang{20}$, and a reported nominal nadir laser footprint of approximately \qty{25}{\centi\meter}. Mean flight-strip spacing was \qty{550}{\meter}, with an overlap of approximately \qty{200}{\meter}. The delivered raw ALS point cloud had a mean density of \qty{31.4}{points\per\square\meter} in forested areas.

To correct for residual positioning errors of the RTK-GNSS plot centers, field plot locations and tree maps derived from PLS were co-registered to the ULS data by matching local PLS-derived canopy height models (CHMs) with the ULS reference CHM. The plot centers exhibited a mean displacement error of \qty{1.31}{\meter}. Additionally, a systematic rotational offset of approximately \ang{14} between local PLS scan frames and the ULS reference frame was identified and manually corrected for each plot.

Because the target population was defined strictly as forest area, circular plot supports were intersected with forest stand boundaries before calculating plot-level GSV and LiDAR metrics. This ensured that response and predictor variables referred to identical spatial domains as the population-level prediction grid. For plots partially overlapping non-forest areas, only the forested fraction of the plot area was retained, and plot-level quantities were expressed per hectare of forest area. Plot-level GSV ranged from 6.7 to 635.7 \unit{\meter\cubed\per\hectare}, with a mean of \qty{321.9}{\meter\cubed\per\hectare} (SD = 124.2).

Two plots were located in a stand that underwent intensive thinning between the ALS (March 2023) and ULS (March 2026) acquisitions. These plots appeared as prominent outliers in the plot-level relationship between ALS- and ULS-derived CHM metrics. Retaining these plots would have severely skewed model comparisons due to high leverage driven by temporal land-cover changes rather than sensor-inherent differences. These two plots were therefore excluded from model fitting and all model-based predictions. However, they were retained for design-based domain estimation (described in Section~\ref{sec:pop_inference}) to preserve the integrity of the probabilistic sampling design.

CHMs were generated from both ULS and ALS point clouds using the R-package \texttt{lidR} \citep{r_lidr_1, r_lidr_2}. These CHMs were aggregated to \qty{3}{\meter} and \qty{35}{\meter} spatial resolutions by computing pixel-wise mean height values. In addition, distributional LiDAR metrics were computed directly from the point clouds at both resolutions. These included the proportion of returns in relative and absolute height bins, as well as measures of vertical point distribution variance. For pixels partially crossing forest boundaries, metrics were calculated exclusively using LiDAR returns contained within the forest mask. Prior to model fitting, all predictor variables were standardized to a zero mean and unit variance separately for each combination of sensor source and spatial resolution.

To mitigate variance inflation caused by multicollinearity and to maintain model interpretability, a parsimonious set of four ecologically meaningful LiDAR metrics was selected manually (Table~\ref{tab:lid_metrics}). Rather than using algorithmic selection or data reduction techniques such as principal component regression \citep{moser_methods_2017, junttila_bayesian_2017}, manual selection ensured a consistent, transparent model structure across all data sources and spatial resolutions while capturing key dimensions of canopy height, canopy density, and vertical structure \citep{lefsky_patterns_2005, li_comparison_2008, white_model_2017}.

To evaluate the added value of calculating comprehensive structural metrics versus relying on basic elevation data, reduced baseline models were fitted using only CHM mean height as a predictor. Basic CHM layers are often readily available from public surveys or photogrammetric processing of low-cost RGB drone imagery. These baseline models thus served to quantify the gain in predictive performance achieved by acquiring dense point clouds and processing the full suite of LiDAR metrics.

\begingroup
\begin{table}
    \centering
\caption{Description of selected predictors.}
\label{tab:lid_metrics}
\renewcommand{\arraystretch}{1.25}
\small
    \begin{tabular}[t]{>{\raggedright\arraybackslash}p{0.15\linewidth}>{\raggedright\arraybackslash}p{0.5\linewidth}>{\raggedright\arraybackslash}p{0.25\linewidth}}
    \toprule
        \textbf{Variable} & \textbf{Description} & \textbf{Interpretation}\\
        \midrule
        \textit{chm} & Mean height of CHM.& Overall forest height\\
        \textit{pct\_above2} & Proportion of returns above \qty{2}{\meter} relative to all returns (including ground returns). & Canopy density\\
        \textit{pct\_10\_20} & Proportion of returns between \qty{10}{\meter} and \qty{20}{\meter} relative to all returns above \qty{2}{\meter}. & Understorey density\\
        \textit{cv\_z} & Coefficient of variation of return heights (sd/mean) for all returns above \qty{2}{\meter}.& Vertical heterogeneity\\
        \bottomrule
    \end{tabular}

\end{table}
\endgroup 
\subsection{Model definition}

The change-of-support (COS) model employed in this study is based on the spatial framework developed by \cite{zhang_bayesian_2024} and implemented in the R package \texttt{cosTapered} \citep{r_costapered}. This framework is specifically tailored for forest inventory settings where the target field response and remotely sensed auxiliary variables are observed across incompatible spatial supports. Here, spatial support means the physical area to which a measurement refers: While LiDAR-derived predictors can be summarized on a fine prediction grid, field response variables are measured on fixed-area circular plots, and final inferential targets represent irregular stand polygons. Rather than projecting all data sources onto a single arbitrary common support prior to modeling, the model estimates an unobserved fine-resolution map (latent response surface) of GSV. Plot and stand values are then obtained by averaging the relevant parts of that map over their respective boundaries. Because GSV is expressed as a density (\unit{\cubic\meter\per\hectare}), areal averaging produces mean GSV for a plot or stand. Stand total volume, if required, is obtained by multiplying mean GSV by stand area. The Bayesian framework allows the uncertainty in the model parameters and the spatial variation to be accounted for jointly, providing both estimates and uncertainty intervals for each stand.

The central concept assumes that an observed plot-level value represents the area-weighted average of the latent, fine-scale response surface over the plot footprint. Similarly, a stand-level value represents the area-weighted average of the same surface over the stand polygon. Thus, forest inventory plots and management units are treated symmetrically as aggregated realizations of the underlying continuous fine-scale process.

\subsubsection{Fine-scale process model}
The latent model on the fine grid describes the response for each raster cell as a function of the auxiliary variables, spatially structured residual variation, and an independent error term. Following the notation of \cite{zhang_bayesian_2024}, let $A_i$ ($i=1,\ldots,n_a$) denote fine grid cells, let $B_l$ ($l=1,\ldots,n_b$) denote observed areal units (sample plots), and $U_k$ ($k=1,\ldots,n_u$) denote target prediction units (e.g.\,, forest stands or holdout polygons).

Let $\bsx(A_i)$ be the $p$-dimensional predictor vector for fine pixel $A_i$, including the intercept if desired. Stacking these vectors yields the $n_a \times p$ design matrix
\begin{equation}
\bm{X} =
\begin{bmatrix}
\bsx(A_1)^\top\\
\vdots\\
\bsx(A_{n_a})^\top
\end{bmatrix}.
\label{eq:X}
\end{equation}
At the fine support, the full Gaussian process (GP) model is formulated as:
\begin{equation}
y(A_i)
=
\bsx(A_i)^\top\bsb
+
\omega(A_i)
+
\epsilon(A_i),
\qquad i=1,\ldots,n_a,
\label{eq:fine_model}
\end{equation}
where $y(A_i)$ represents the latent fine-scale response density, $\bsx(A_i)^\top\bsb$ is the fixed-effects regression component, $\omega(A_i)$ is the spatially structured residual variation, and $\epsilon(A_i)$ represents independent fine-scale noise at the finite grid-cell support. The fine-cell responses are not directly observed; they are latent quantities used to reconcile the differing plot, grid, and stand supports.

Defining the spatial residual and error vectors as
\begin{equation}
\bsw_A = 
\begin{bmatrix}
\omega(A_1) \\
\vdots \\
\omega(A_{n_a})
\end{bmatrix},
\qquad
\bse_A = 
\begin{bmatrix}
\epsilon(A_1) \\
\vdots \\
\epsilon(A_{n_a})
\end{bmatrix},
\label{eq:w_e_vectors}
\end{equation}
the spatial process prior is specified as
\begin{equation}
\bsw_A \sim \mathcal{N}\left(\mathbf{0}, \sigma^2 \mathbf{C}^{\mathrm{tap}}_{\phi,\gamma}\right),
\label{eq:wA_prior}
\end{equation}
with the tapered correlation matrix defined element-wise by
\begin{equation}
\left[\mathbf{C}^{\mathrm{tap}}_{\phi,\gamma}\right]_{ij} = \exp\left\{-\phi d(s_i, s_j)\right\} T_\gamma\left(d(s_i, s_j)\right).
\label{eq:full_exp_cor}
\end{equation}

Here, $s_i$ and $s_j$ denote the centroids of fine-pixels $A_i$ and $A_j$, and $d(s_i,s_j)$ represents their Euclidean distance. The parameter $\sigma^2$ controls the partial sill (spatial variance component), $\phi$ dictates the spatial decay rate of the exponential correlation, and $\gamma$ represents the tapering range. The compact taper $T_\gamma(\cdot)$ evaluates to zero for distances $d(s_i,s_j) \ge \gamma$, thereby setting residual covariance to zero beyond this distance. This approximation significantly accelerates computation because the compact support makes the covariance matrices sparse (meaning that most matrix entries are zero) \citep{zhang_bayesian_2024}. For the exponential component, the effective spatial range is defined as the distance at which spatial correlation drops to $0.05$, calculated as $-\log(0.05)/\phi \approx 3/\phi$; note that this effective range is conceptually distinct from the hard taper range $\gamma$.  In all spatial models, the Wendland taper was used with $\gamma=\qty{1000}{\meter}$. This spatial process allows nearby locations to have similar unexplained GSV values after accounting for LiDAR predictors, while similarity declines with distance.

The independent noise component (nugget) is modeled as
\begin{equation}
\bse_A \sim \mathcal{N}\left(\bm{0},\tau^2\bm{I}_{n_a}\right) ,
\label{eq:eA_prior}
\end{equation}
where $\tau^2$ denotes the nugget variance defined explicitly at the discrete grid-cell support rather than as an integrated continuous white-noise process. $\bse_A$ represents local, unstructured variation at the grid-cell scale.

\subsubsection{Observed areal model}
The central step of the change-of-support approach is to transfer the latent response modeled on the fine grid to the spatial support of the observed sample plots. This is achieved through an area-weighted aggregation of the fine-grid cells, thereby explicitly accounting for the different spatial supports of the auxiliary data and observations. Each observed plot $B_l$ is linked to the fine predictor grid through an area-based weighting factor,
\begin{equation}
h_{li}
=
\frac{|B_l \cap A_i|}{|B_l|},
\label{eq:hli}
\end{equation}
where $|\cdot|$ denotes surface area. Thus, $h_{li}$ represents the proportion of plot $B_l$'s area contributed by fine pixel $A_i$. For example, if a grid cell contributes 10\% of a plot’s area, its fine-scale value receives weight ($h_{li} = 0.10$) in the plot average. 

Let $\bm{H}_{BA}$ be the $n_b \times n_a$ aggregation matrix with entries $h_{li}$, where each row corresponds to an observed plot $B_l$ and each column corresponds to a fine grid cell $A_i$. Each row contains the area-based weights used to aggregate fine-scale values to that plot. By construction, rows of $\bm{H}_{BA}$ sum to unity provided that the fine grid fully covers each observed support. In this global aggregation matrix $\mathbf{H}_{BA}$, $h_{li}=0$ for all fine pixels that do not intersect plot $B_l$. Consequently, $\mathbf{H}_{BA}$ remains highly sparse even when the fine grid completely covers the study region.

The observed response on plot $B_l$ is defined as the area-weighted average of the underlying latent process:
\begin{equation}
y(B_l)
\approx
\sum_{i=1}^{n_a} h_{li} y(A_i).
\label{eq:areal_average}
\end{equation}
Stacking observed plot responses into the $n_b \times 1$ vector $\bsy_B$ yields
\begin{equation}
\bsy_B
=
\bm{H}_{BA}\bm{X}\bsb
+
\bm{H}_{BA}\bsw_A
+
\bm{H}_{BA}\bse_A.
\label{eq:yB_fine_stack}
\end{equation}
Defining the areal spatial random effects as
\begin{equation}
\bsw_B = \bm{H}_{BA}\bsw_A.
\label{eq:wB_def}
\end{equation}
and leveraging the Gaussian property of $\boldsymbol{\omega}_A$, we obtain
\begin{equation}
\bsw_B \sim N(\bm{0},\sigma^2\bm{C}^{\mathrm{tap}}_B(\phi,\gamma)),
\label{eq:wB_prior}
\end{equation}
where the aggregated correlation matrix is defined as
\begin{equation}
\bm{C}^{\mathrm{tap}}_B(\phi,\gamma)
=
\bm{H}_{BA}\bm{C}^{\mathrm{tap}}_{\phi,\gamma}\bm{H}_{BA}^\top.
\label{eq:CB_full}
\end{equation}

Similarly, the aggregated measurement error transforms to
\begin{equation}
\bm{H}_{BA}\bse_A
\sim
N(\bm{0},\tau^2 \bm{D}_h),
\qquad
\bm{D}_h = \bm{H}_{BA}\bm{H}_{BA}^\top.
\label{eq:Dh}
\end{equation}
If the observed supports do not share any fine grid cells, $\bm{D}_h$ reduces to a diagonal matrix. If sample plots overlap at the fine-cell level, $\bm{D}h$ contains off-diagonal elements. Preserving $\bm{D}_h=\bm{H}_{BA}\bm{H}_{BA}^\top$ in the observation-level covariance structure accommodates both scenarios without requiring a diagonal nugget approximation. Although the nugget term is independent among fine grid cells, plot responses average these cell-level deviations across multiple cells, thereby reducing nugget variance at the plot level. When two plots include one or more of the same fine cells, they inherit some of the same cell-level deviations, producing correlated nugget terms.

The full conditional model for the observed support is therefore expressed as
\begin{equation}
\bsy_B \mid \bsb,\bsw_B,\tau^2
\sim
N\left(
\bm{H}_{BA}\bm{X}\bsb+\bsw_B,\,
\tau^2\bm{D}_h
\right),
\label{eq:yB_conditional}
\end{equation}
with
\begin{equation}
\bsw_B\mid \sigma^2,\phi,\gamma
\sim
N\left(\bm{0},\sigma^2\bm{C}^{\mathrm{tap}}_B(\phi,\gamma)\right).
\label{eq:wB_conditional}
\end{equation}

This marginal formulation avoids direct sampling of the high-dimensional latent vector $\bsw_A$ during MCMC estimation. Instead, model fitting is performed on the much lower-dimensional areal space of $\bsw_B$, drastically reducing computational demands as demonstrated by \cite{zhang_bayesian_2024}.

\subsubsection{Prior specification}
Prior distributions for all model parameters were specified as
\begin{align}
\bsb &\sim N(\bm{\mu}_{\beta},\bm{V}_{\beta}), \label{eq:beta_prior}\\
\tau_B^2 &\sim \mathrm{IG}(t_{\tau},s_{\tau}), \label{eq:tau_prior}\\
\sigma^2 &\sim \mathrm{IG}(t_{\sigma},s_{\sigma}), \label{eq:sigma_prior}\\
\phi &\sim \mathrm{Unif}(a_{\phi},b_{\phi}), \label{eq:phi_prior}
\end{align}
where $\mathrm{IG}(t, s)$ denotes the inverse-gamma distribution parameterized by shape $t$ and scale $s$, with probability density function
\begin{equation}
p(x \mid t, s)
=
\frac{s^t}{\Gamma(t)}
x^{-(t+1)}
\exp\{-s/x\},
\qquad x>0.
\label{eq:ig_density}
\end{equation}
Although the core model is formulated in terms of the fine-scale nugget variance $\tau^2$, setting a prior directly on $\tau^2$ can be unintuitive because the observed plot data experience the measurement error only after spatial support aggregation. Following the reparameterization strategy implemented in  \texttt{cosTapered}, we specified the nugget prior on the average observed-support nugget variance $\tau_B^2$, defined as
\begin{equation*}
\tau_B^2 = \tau^2 \bar{d}_h,
\qquad
\bar{d}_h =
\frac{1}{n_b}\sum_{l=1}^{n_b} D_h(l,l).
\label{eq:tauB_reparam}
\end{equation*}
Here, $\tau_B^2$ represents the average diagonal contribution of the nugget error to the observed plot-level covariance, given that the aggregated nugget covariance matrix is $\tau^2\bm{D}_h$ with average diagonal element $\tau^2\bar{d}_h$. This conversion allows exact back-transformation to the fine-scale nugget variance:
\begin{equation*}
\tau^2 = \frac{\tau_B^2}{\bar{d}_h}.
\label{eq:tau_backtransform}
\end{equation*}
This reparameterization leaves the underlying statistical model unchanged while providing a substantially more interpretable target for prior elicitation. Note that $\tau_B^2$ represents the average plot-level nugget variance across all plots; for a specific plot $l$, the exact plot-level nugget variance remains $(\tau_B^2/\bar{d}_h)D_h(l,l)$.

Defining the vector of variance and spatial decay parameters as
\begin{equation}
\bstheta = (\tau^2,\sigma^2,\phi),
\label{eq:theta_def}
\end{equation}
the MCMC sampler operates on the equivalent working parameterization $\bstheta_B=(\tau_B^2,\sigma^2,\phi)$, transforming each sampled $\tau_B^2$ draw back to $\tau^2$ prior to evaluating the likelihood functions and generating posterior predictive realizations.

Weakly informative priors were assigned to all parameters to regularize estimation by excluding implausible parameter values while allowing the observed inventory data to dominate posterior estimates. Because continuous predictor variables were standardized prior to model fitting, regression coefficients describe the expected change in GSV (in \unit{\cubic\meter\per\hectare}) per one-standard-deviation change in the corresponding LiDAR metric. The intercept was assigned a weakly informative normal prior centered on the sample mean of observed GSV, whereas regression coefficients were assigned zero-centered normal priors with variance parameters informed by preliminary exploratory linear regressions. Inverse-gamma priors for $\tau_B^2$ and $\sigma^2$ were chosen to be weakly informative over plausible variance ranges. For the spatial decay parameter $\phi$, the uniform prior bounds were informed by empirical variogram analysis of the field response and preliminary spatial linear mixed models fitted using the R package \texttt{stLMM} \citep{r_stlmm}; these bounds correspond to exponential effective ranges of approximately \qtyrange{30}{1000}{\meter}. Complete hyperparameter specifications for all fitted models are summarized in Table~\ref{tab:priors}.

\begin{table}
    \centering
    \small
    \begin{tabular}{cccc}\toprule
         Parameter & Intercept-only & CHM-only & All predictors\\\midrule
         $\beta_0$ & $\mathcal{N}(324.9, 125^2)$ & $\mathcal{N}(324.9, 125^2)$& $\mathcal{N}(324.9, 125^2)$\\
         $\beta_n$ & -- & $\mathcal{N}(0, 125^2)$ & $\mathcal{N}(0, 75^2)$\\
         $\tau_B^2$& $\mathrm{IG}(2, 7000)$ &  $\mathrm{IG}(2, 4000)$ &  $\mathrm{IG}(2, 3000)$\\
         $\sigma^2$ & $\mathrm{IG}(2, 7000)$ & $\mathrm{IG}(2, 4000)$ & $\mathrm{IG}(2, 3000)$\\
         $\phi$& \multicolumn{3}{c}{$\mathrm{Unif}(3/1000, 3/30)$}\\ \bottomrule
    \end{tabular}
    \caption{Summary of priors used for model fitting. Priors on $\sigma^2$ and $\phi$ only apply to spatial models. All other priors are identical for non-spatial variants. Priors for both resolution model variants are also identical.}
    \label{tab:priors}
\end{table} 
\subsubsection{Posterior sampling strategy}
Model parameters were estimated using the Markov chain Monte Carlo (MCMC) sampling scheme implemented in the R package \texttt{cosTapered} \citep{r_costapered}. The sampler updates the spatial covariance and variance parameters from a collapsed posterior distribution in which the regression coefficients $\boldsymbol{\beta}$ and spatial random effects $\boldsymbol{\omega}_B$ are integrated out analytically. Subsequently, posterior draws for $\boldsymbol{\beta}$, $\boldsymbol{\omega}_B$, and the observed-support latent mean process are recovered conditionally. Detailed algorithmic derivations, including tapered covariance matrix constructions and exact conditional recovery distributions, are provided by \cite{zhang_bayesian_2024} and in the \texttt{cosTapered} package documentation \citep{r_costapered}.

For each model specification, three independent MCMC chains were initialized from dispersed starting values and run for 5,000 iterations per chain. Posterior summaries of covariance and variance parameters were computed from pooled post-burn-in draws after discarding the first 1,000 iterations of each chain. Regression coefficients, observed-support spatial effects, and stand-level predictions were then recovered from a random subset of 600 post-burn-in draws to limit storage and computation. MCMC convergence was evaluated using the Gelman-Rubin convergence diagnostic $\hat{R}$ alongside visual inspection of trace plots for $\tau_B^2$, $\sigma^2$, and $\phi$. Chains were considered fully converged when $\hat{R} < 1.01$ and trace plots exhibited stable mixing with good chain overlap. The retained posterior draws represent plausible parameter values and spatial predictions after accounting for both data uncertainty and spatial dependence.

\subsection{Model evaluation}
The Bayesian $R^2$ introduced by \cite{gelman_r-squared_2019} was computed as an initial descriptive metric of in-sample model fit. It is defined as the variance of the fitted values divided by the sum of the variance of the fitted values and the expected residual variance. For the spatial COS models, a conditional Bayesian $R^2$ was calculated at the observed plot support as
\begin{equation}
R^2 =
\frac{
\mathrm{Var}\left(\bm{H}_{BA}\bm{X}\boldsymbol{\beta} + \bsw_B\right)
}{
\mathrm{Var}\left(\bm{H}_{BA}\bm{X}\boldsymbol{\beta} + \bsw_B\right) + \tau_B^2
}
\label{eq:bayesian_r2}
\end{equation}
Because the fitted spatial effect $\bsw_B$ is included in the numerator, this is a conditional $R^2$ because it describes variation represented jointly by the LiDAR predictors and the fitted spatial pattern, rather than the contribution of LiDAR predictors alone. Here, $\tau_B^2$ represents the average residual nugget variance at plot support.
This calculation was performed for each retained posterior draw, and results were summarized using the posterior median and \qty{95}{\percent} credible interval. 
Because $R^2$ characterizes in-sample explained variation without penalizing model complexity or assessing predictive generalization, it was used strictly as a descriptive measure of fit. 

Out-of-sample predictive performance was evaluated using $10$-fold cross-validation. Sample plots were randomly partitioned into 10 folds of approximately equal size ($14$--$15$ plots per fold). For each fold, models were fitted to the remaining nine training folds and used to generate posterior predictive distributions for the held-out validation observations. Point prediction accuracy across all held-out plots was summarized using bias, root mean squared prediction error (RMSPE), and mean absolute error (MAE), calculated from the differences between observed values and posterior predictive means.

Probabilistic predictive performance and uncertainty calibration were evaluated using the Continuous Ranked Probability Score (CRPS; \cite{gneiting_strictly_2007}), empirical coverage of the \qty{95}{\percent} posterior predictive intervals, and mean prediction interval width (MPIW). For each held-out observation $i$, the CRPS was computed directly from posterior predictive samples using the empirical scoring rule
\begin{equation}
\mathrm{CRPS}(F_i,y_i)
=
\mathbb{E}_{F_i}|Y_i-y_i|
-
\frac{1}{2}\mathbb{E}_{F_i}|Y_i-Y_i'|,
\label{eq:crps}
\end{equation}
where $F_i$ represents cumulative posterior predictive distribution for held-out observation $i$, $y_i$ is the observed field response, and $Y_i, Y_i' \sim F_i$ are independent posterior draws. Lower CRPS values indicate superior probabilistic predictions by rewarding both high predictive accuracy and appropriate scale of uncertainty. CRPS can thus be interpreted as an accuracy score for an entire predictive distribution, rather than for its mean prediction alone.

Empirical coverage was computed as the proportion of held-out observations falling within their corresponding \qty{95}{\percent} posterior predictive intervals. MPIW was calculated as the mean width of these prediction intervals across all validation plots. Together, empirical coverage measures the calibration of predictive uncertainty, while MPIW characterizes the sharpness (precision) of the predictive distributions. A well-calibrated (\qty{95}{\percent}) prediction interval should contain approximately (\qty{95}{\percent}) of held-out observations. Coverage should be interpreted jointly with MPIW, because overly wide intervals can achieve high coverage while providing little practical precision.

\subsection{Fine-scale recovery and areal prediction}
For spatial mapping and stand-level inference, the primary target is the latent fine-scale mean surface, defined as
\begin{equation}
\bm{\eta}_A
=
\bm{X}\bsb+\bsw_A.
\label{eq:eta_A}
\end{equation}
By excluding the fine-scale nugget variability $\bse_A$, $\boldsymbol{\eta}_A$ represents the underlying, smoothed spatial forest response density at the pixel grid resolution. This means that the map represents expected or underlying GSV, rather than the value that might be observed in one particular fine grid cell after local random variation.

To predict spatial averages over arbitrary target polygons $U_k$ ($k = 1, \ldots, n_u$), such as forest management stands or inventory strata, we construct an area-based projection matrix $\mathbf{H}_{UA}$ of dimension $n_u \times n_a$. The elements $g_{ki}$ represent the proportion of target unit $U_k$'s area covered by fine pixel $A_i$:
\begin{equation}
g_{ki}
=
\frac{|U_k\cap A_i|}{|U_k|}.
\label{eq:gki}
\end{equation}
The latent mean vector for all target prediction units, $\boldsymbol{\eta}_U$, is then obtained via linear aggregation as
\begin{equation}
\bm{\eta}_U
=
\bm{H}_{UA}\bm{\eta}_A.
\label{eq:eta_U}
\end{equation}

Thus, stand predictions are obtained using the same area-weighted averaging principle used for field plots. The resulting values represent mean GSV density within each stand; predicted total volume can be obtained by multiplying mean GSV by stand area.

Stand- and population-level predictions targeted the latent spatial mean surface, excluding fine-scale nugget variability. For each areal prediction unit, posterior uncertainty summaries incorporated marginal uncertainty in the unobserved spatial effect at that support. These summaries therefore describe each target unit considered separately; they are not joint posterior draws over all stands and should not be used directly for totals, rankings, or other summaries requiring cross-stand covariance.

For target units containing incomplete predictor grid cells (e.g.,\,due to localized low point densities in LiDAR acquisitions that prevented predictor calculation), the intersection weights $g_{ki}$ of the remaining valid cells were re-normalized to sum to unity within that unit. This procedure treats the valid predictor cells as representative of the full target unit. Predictions for units with substantial missing predictor coverage should therefore be interpreted with caution.

\subsection{Population-level inference} \label{sec:pop_inference}
To provide a benchmark for evaluating model-based predictions, design-based estimators of population mean GSV were computed using all 151 sample plots. Unlike the COS predictions, which use a statistical model relating GSV to LiDAR and spatial structure, design-based estimates rely on the sampling design and observed field measurements. They provide an independent benchmark that does not require the spatial model to be correct.

The spatial georeferencing of individual trees enabled tree-level Horvitz--Thompson (HT) weighting to account for boundary-overlap effects and varying inclusion probabilities. Specifically, each tree's inclusion probability was calculated as the proportion of the area of a \qty{20}{\meter} radius circle centered on the tree that fell within the forest sampling frame. The resulting tree-level values were aggregated to obtain design-unbiased plot-level GSV estimates. 

Using these plot-level estimates, population-level parameters were derived under several classical sampling designs. First, a design-based estimator assuming systematic sampling (SYS) was calculated. Second, several post-stratification estimators were evaluated, leveraging both existing stand-level inventory records and the 2024 CORINE Dominant Leaf Type \qty{10}{\meter} raster product \citep{eea_dominant_2024}. 
Post-stratification schemes included:
\begin{enumerate}
\item forest composition derived from historical stand-level inventory data (broadleaf vs.\ conifer),
\item forest composition derived from the satellite-based CORINE raster, and
\item stand age classes derived from inventory records, grouped into four categories ($0$--$40$, $41$--$80$, $81$--$120$, and $\ge 121$~years)
\end{enumerate}
Post-stratification improves precision when the strata are related to GSV and their area proportions are known for the full forest population.

Finally, a classic difference-cum-regression estimator was calculated as
\begin{equation}
\bar{y}_R = \bar{y} + b_1(\mu_x - \bar{x}),
\label{eq:regression_estimator}
\end{equation}
where $\bar{y}$ represents the HT estimator of population mean GSV, $\bar{x}$ is the sample mean of the ULS-derived CHM evaluated across all sample plots, $\mu_x$ is the true population mean of the ULS-CHM computed across the entire forest area, and $b_1$ is the ordinary least-squares regression coefficient of GSV on CHM at plot-level.

\section{Results}

\subsection{Model coefficients and spatial parameters}

\begin{figure}
    \centering
    \includegraphics[width=1\linewidth]{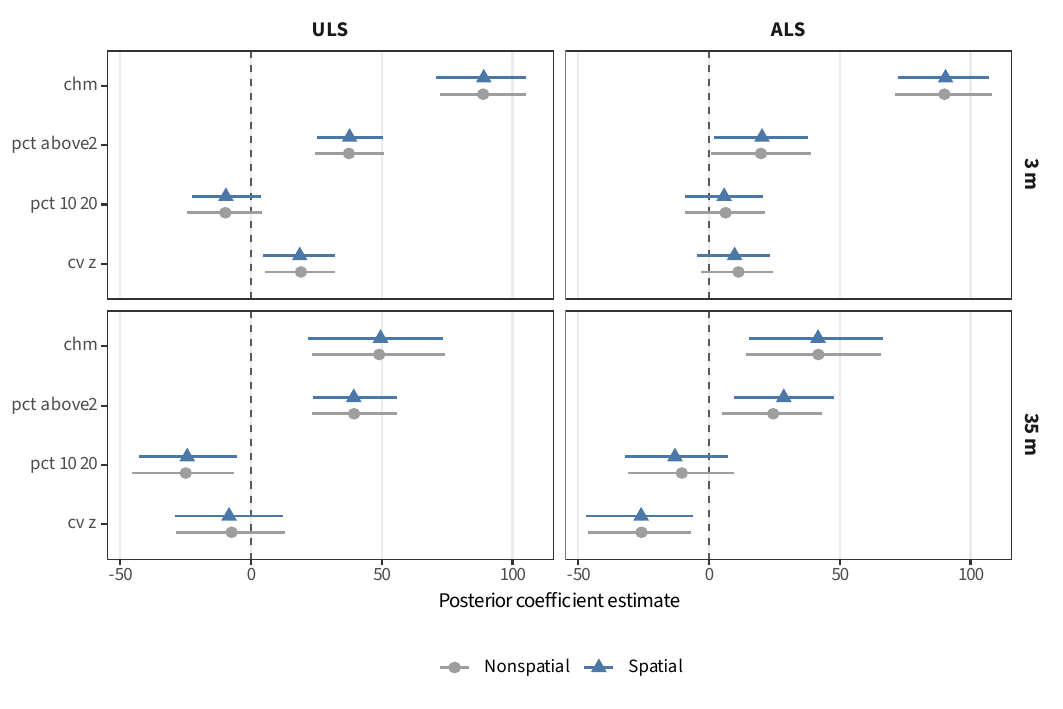}
    \caption{ Posterior parameter estimates. Points represent posterior medians with 95\% credible intervals as lines. Predictors are standardized separately for each data source and resolution.}
    \label{fig:posterior_forest}
\end{figure}

The posterior summaries for the regression coefficients of the full-predictor models are visually summarized in Figure~\ref{fig:posterior_forest} (with detailed numerical values provided in Tables~\ref{tab:pp_full_3_app}--\ref{tab:pp_full_35_app} in Appendix~\ref{sec:appendix_models}). Baseline models (intercept-only and CHM-only variants) serve as benchmarks and are provided in Appendix~\ref{sec:appendix_models} (Tables~\ref{tab:pp_red_3_app}--\ref{tab:pp_red_35_app}), as well as in the subsequent cross-validation analysis.

Mean CHM height and canopy density showed consistent positive associations across all eight full-metrics models, with credible intervals entirely above zero. In contrast, understory density and vertical heterogeneity varied more strongly between data sources and resolutions, both in the direction of the posterior median and in whether their credible intervals included zero. For the \qty{3}{\meter} models, the CHM coefficient was larger than the canopy-density coefficient. At \qty{35}{\meter} resolution, the corresponding credible intervals largely overlapped. Spatial and non-spatial model variants gave broadly similar coefficient estimates, with only minor differences in posterior medians and interval widths.

At the same spatial resolution, the proportion of fine-scale residual variance attributable to the spatial component, $\sigma^2 / (\sigma^2 + \tau^2)$, was slightly larger for the ALS models. The fine-scale nugget variance, $\tau^2$, was several orders of magnitude larger in the \qty{3}{\meter} change-of-support models than in the \qty{35}{\meter} models while $\sigma^2$ was similar, but slightly larger in \qty{35}{\meter} variants. Posterior medians of the effective range were consistently larger for model variants using the ALS parameter set, although the credible intervals were wide and overlapped substantially. For the ALS models, the posterior median effective range exceeded the sample grid spacing of \qty{183}{\meter}, but the uncertainty around this estimate remained high.

The non-spatial variants differed in explained variance as measured by Bayesian $R^2$. For ULS models, $R^2$ was considerably higher than for ALS models at the same resolution, and the \qty{3}{\meter} models also showed higher $R^2$ than the corresponding \qty{35}{\meter} models. These differences were not apparent in the spatial models, for which Bayesian $R^2$ was similar across data sources and resolutions. This suggests that variation not captured by the fixed effects in the non-spatial models may be partly accounted for by the spatial random field. However, because $R^2$ in the spatial models included variation explained by $\omega$, it should not be interpreted solely as covariate-explained variance. Cross-validation therefore provides a more direct basis for comparing predictive performance.

\subsection{Cross-validation performance}
\begin{figure}
    \centering
    \includegraphics[width=1\linewidth]{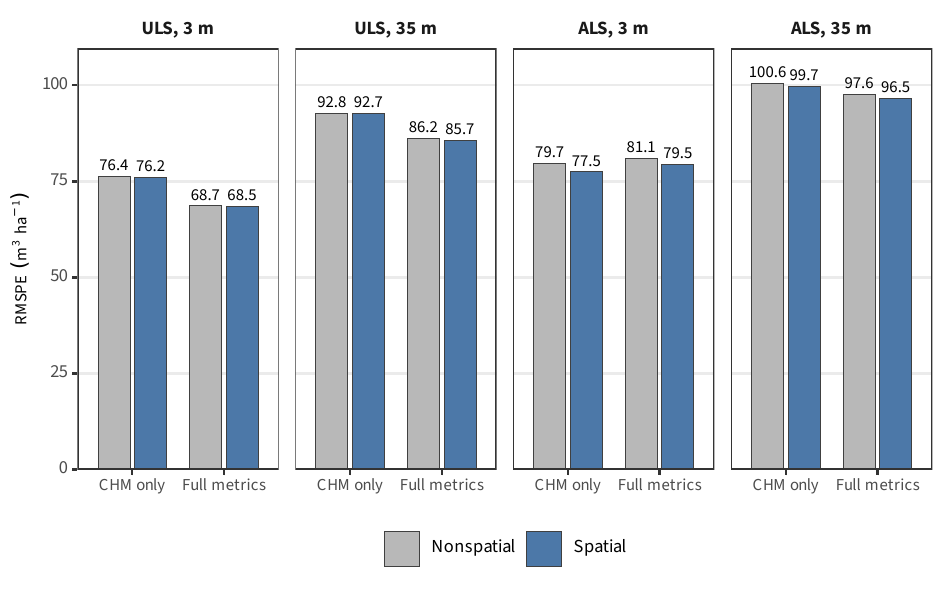}
    \caption{Root mean squared predictive error (RMSPE) evaluated by 10-fold cross-validation.}
    \label{fig:cv_columns}
\end{figure}

The results of the 10-fold CV are presented in Figure~\ref{fig:cv_columns} and Table~\ref{tab:cv_metrics}. Overall, RMSPE, MAE, and CRPS largely agree on the ordering of the models by predictive error. Including the baseline models highlights the incremental value of adding LiDAR structural predictors. Relative to the intercept-only baseline ($\text{RMSPE} = \qty{122.9}{\cubic\metre\per\hectare}$ at \qty{3}{\meter} resolution), incorporating CHM height substantially reduces predictive error across all data sources and resolutions.

In direct comparisons, all \qty{3}{\meter} COS model variants performed considerably better than their corresponding \qty{35}{\meter} counterparts. Reduced models with mean CHM height as the only linear predictor performed similarly for ULS and ALS, with a slight advantage for drone-based LiDAR (\text{RMSPE} = \qty{76.4}{\cubic\metre\per\hectare} for the ULS non-spatial model versus \qty{79.7}{\cubic\metre\per\hectare} for the ALS non-spatial model at \qty{3}{\meter} resolution). However, when additional distributional predictors were included, the ULS models showed considerably larger improvements. Specifically, RMSPE was reduced by 10.1\% for both the non-spatial and spatial variants at \qty{3}{\meter} resolution, and by 7.1\% and 7.6\%, respectively, at \qty{35}{\meter} resolution, relative to the corresponding CHM-only models. For ALS models, adding distributional predictors improved predictive performance only at \qty{35}{\meter} resolution, yielding smaller RMSPE reductions of 3.0\% and 3.2\% for the non-spatial and spatial variants, respectively. At \qty{3}{\meter} resolution, adding distributional metrics to ALS increased error for both model variants; for example, non-spatial RMSPE increased from \qty{79.7}{\cubic\metre\per\hectare} to \qty{81.1}{\cubic\metre\per\hectare}.

Spatial variants generally performed similarly to, or slightly better than, their non-spatial counterparts. Although the improvements were generally small, gains from including the spatial process were more pronounced for ALS than for ULS. The largest improvement was observed for the ALS \qty{3}{\meter} models, where inclusion of a spatial process reduced RMSPE by 2.0\% for the full predictor set and by 2.8\% for the CHM-only model.

Bias was low overall and consistently negative across all models, indicating a slight tendency toward underprediction. The largest absolute bias occurred in the ALS \qty{35}{\meter} full-metrics non-spatial model (\qty{-7.7}{\cubic\meter\per\hectare}), whereas the smallest absolute bias was observed in the ULS \qty{3}{\meter} full-metrics spatial model (\qty{-0.3}{\cubic\meter\per\hectare}). Adding distributional predictors reduced absolute bias for several model variants, particularly the spatial ULS models, but increased it slightly in some non-spatial models.

Coverage of the 95\% prediction intervals remained consistent across models, ranging from 93.3\% to 96.6\%. Mean prediction interval width (MPIW) differed more clearly between model variants and followed a broadly similar ordering to the error metrics. The ULS \qty{3}{\meter} full-metrics non-spatial model achieved the narrowest intervals (\qty{266.7}{\cubic\meter\per\hectare}), whereas the intercept-only baseline ($\approx$ \qty{480}{\cubic\meter\per\hectare}) and ALS \qty{35}{\meter} models (\qtyrange{373.2}{392.7}{\cubic\meter\per\hectare}) produced the widest intervals. This pattern indicates that models with lower predictive error generally yielded sharper predictions. In absolute terms, interval widths remained wide for individual plot-level predictions when compared to the observed response, ranging from \qtyrange{266.7}{380.2}{\cubic\meter\per\hectare} for non-spatial models with all predictors. This behaviour is expected for small plot supports affected by fine-scale heterogeneity and support mismatch. Stand-level aggregation over larger units is evaluated in the following Section~\ref{sec:stand_pred}.

\begin{table}
\centering
\caption{Cross-validation performance of the fitted models. RMSPE, MAE, bias, CRPS and MPIW are expressed in \unit{\cubic\metre\per\hectare}. Coverage is expressed as a percentage.}
\label{tab:cv_metrics}
\centering
\small
\begin{tabular}[t]{llrrrrrl}
\toprule
\textbf{Model} & \textbf{Structure} & \textbf{RMSPE} & \textbf{MAE} & \textbf{Bias} & \textbf{CRPS} & \textbf{Coverage (95\%)} & \textbf{MPIW}\\
\midrule
\addlinespace[0.3em]
\multicolumn{8}{l}{\textbf{Intercept only}}\\
\hspace{1em}\qty{3}{\metre} & Spatial & 122.9 & 98.5 & -0.0 & 69.8 & 96.0 & 480.0\\
\hspace{1em}\qty{35}{\metre} & Spatial & 123.4 & 98.9 & -3.5 & 70.1 & 94.0 & 480.1\\
\addlinespace[0.3em]
\multicolumn{8}{l}{\textbf{ULS, 3 m}}\\
\hspace{1em}CHM only & Spatial & 76.2 & 57.8 & -1.2 & 41.6 & 95.3 & 300.1\\
\hspace{1em}CHM only & Nonspatial & 76.4 & 58.0 & -2.1 & 41.7 & 94.6 & 297.1\\
\hspace{1em}Full metrics & Spatial & 68.5 & 51.6 & -0.3 & 37.6 & 94.6 & 269.6\\
\hspace{1em}Full metrics & Nonspatial & 68.7 & 51.6 & -2.3 & 37.5 & 93.3 & 266.7\\
\addlinespace[0.3em]
\multicolumn{8}{l}{\textbf{ULS, 35 m}}\\
\hspace{1em}CHM only & Spatial & 92.7 & 73.4 & -4.5 & 51.5 & 96.0 & 361.0\\
\hspace{1em}CHM only & Nonspatial & 92.8 & 72.8 & -6.1 & 51.5 & 96.6 & 365.5\\
\hspace{1em}Full metrics & Spatial & 85.7 & 66.6 & -2.8 & 47.7 & 95.3 & 338.7\\
\hspace{1em}Full metrics & Nonspatial & 86.2 & 66.4 & -6.0 & 47.8 & 94.0 & 337.9\\
\addlinespace[0.3em]
\multicolumn{8}{l}{\textbf{ALS, 3 m}}\\
\hspace{1em}CHM only & Spatial & 77.5 & 58.5 & -2.1 & 42.4 & 95.3 & 306.0\\
\hspace{1em}CHM only & Nonspatial & 79.7 & 60.0 & -2.8 & 43.5 & 94.0 & 310.2\\
\hspace{1em}Full metrics & Spatial & 79.5 & 60.9 & -1.1 & 43.9 & 96.0 & 310.9\\
\hspace{1em}Full metrics & Nonspatial & 81.1 & 61.5 & -2.9 & 44.4 & 94.6 & 312.6\\
\addlinespace[0.3em]
\multicolumn{8}{l}{\textbf{ALS, 35 m}}\\
\hspace{1em}CHM only & Spatial & 99.7 & 78.5 & -5.3 & 55.4 & 96.0 & 389.7\\
\hspace{1em}CHM only & Nonspatial & 100.6 & 78.7 & -6.9 & 55.8 & 94.0 & 392.7\\
\hspace{1em}Full metrics & Spatial & 96.5 & 77.0 & -4.5 & 54.1 & 95.3 & 373.2\\
\hspace{1em}Full metrics & Nonspatial & 97.6 & 77.0 & -7.7 & 54.2 & 94.0 & 380.2\\
\bottomrule
\end{tabular}
\end{table}
 
\subsection{Stand-level predictions} \label{sec:stand_pred}
\begin{figure}
    \centering
    \includegraphics[width=0.95\linewidth]{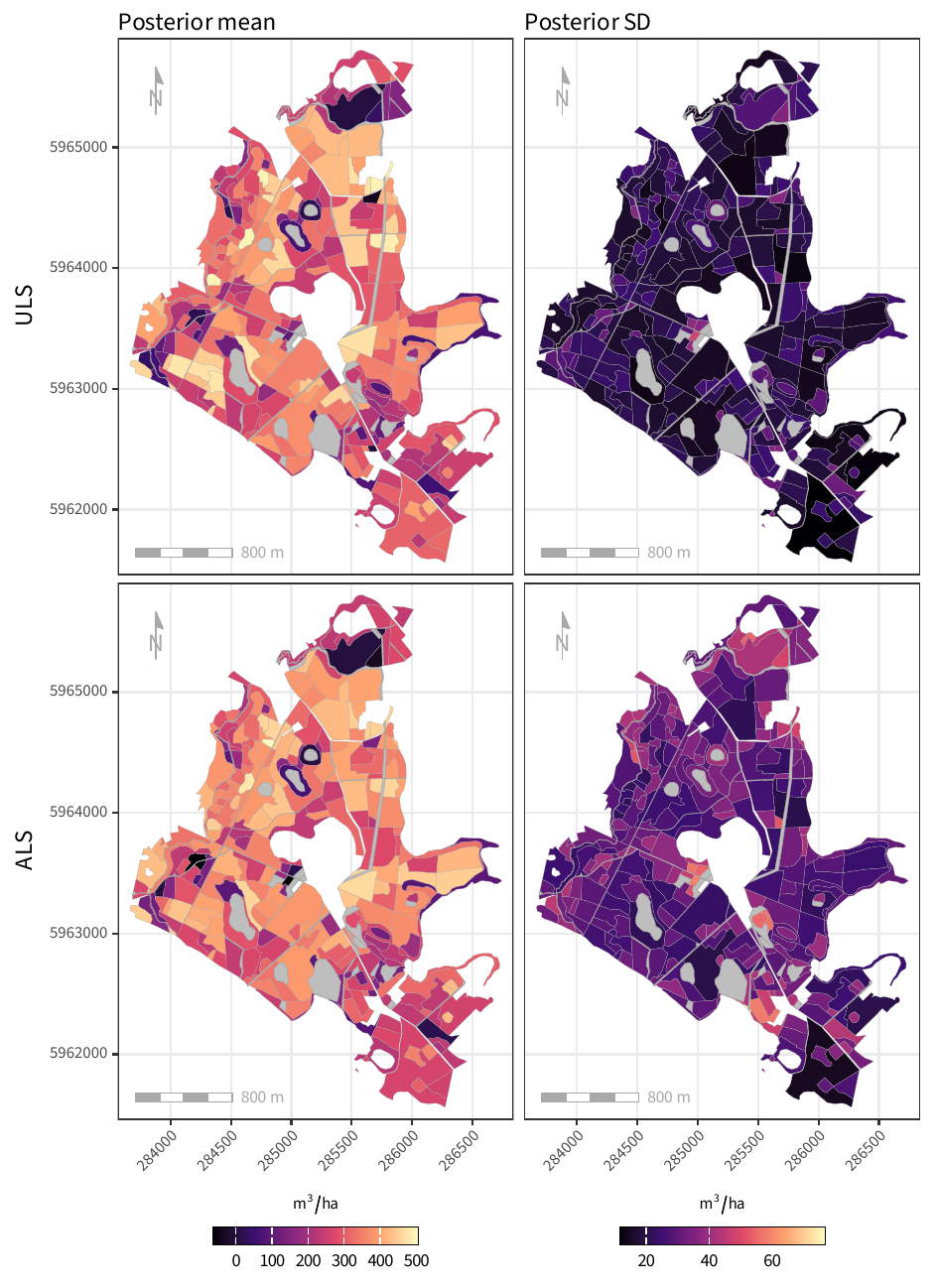}
    \caption{Stand-level posterior mean and posterior standard deviation for latent GSV using spatial models with full predictor sets at \qty{3}{\meter} resolution, comparing ULS and ALS data sources.}
    \label{fig:predictions}
\end{figure}

\begin{figure}
    \centering
    \includegraphics[width=1\linewidth]{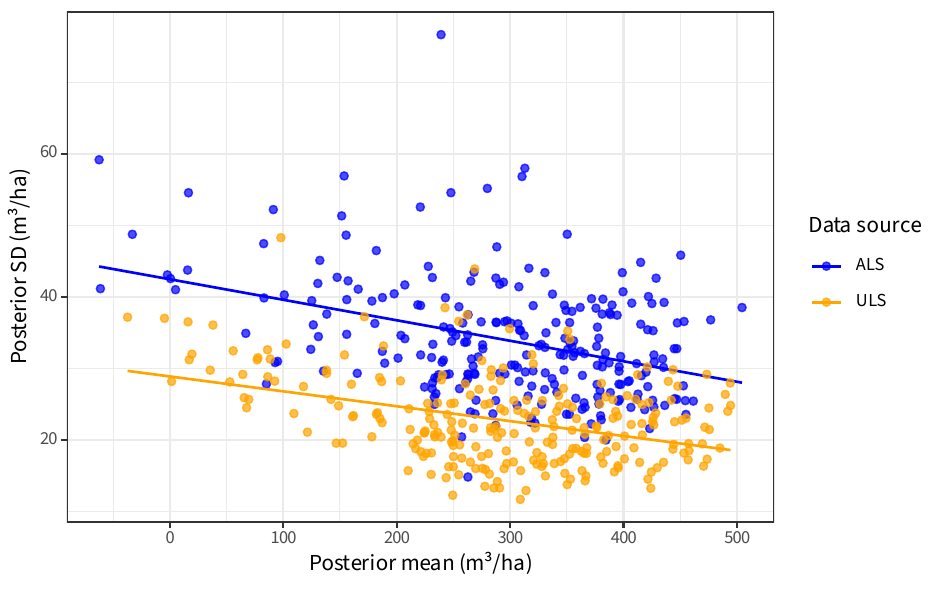}
    \caption{Relationship between stand-level posterior GSV (posterior mean) and prediction uncertainty (posterior SD) for spatial models at \qty{3}{\meter} resolution across ULS and ALS platforms. Solid lines show ordinary least-squares fits per platform.}
    \label{fig:mean_error}
\end{figure}

Stand-level predictions for forest stands using the \qty{3}{\metre} spatial models with full predictor sets are illustrated in Figure~\ref{fig:predictions} for both ALS and ULS data sources. Neither model reveals prominent large-scale spatial patterns in posterior standard deviation. However, overall uncertainty is noticeably higher for the ALS model, with posterior SD ranging from \qtyrange{14.8}{76.6}{\cubic\meter\per\hectare}, compared to \qtyrange{11.7}{48.3}{\cubic\meter\per\hectare} for the ULS model. As depicted in Figure~\ref{fig:mean_error}, stand-level uncertainty is elevated in low-volume stands for both platforms. It decreases as predicted GSV increases, although there is considerable scatter around the fitted line. 

\begin{figure}
    \centering
    \includegraphics[width=1\linewidth]{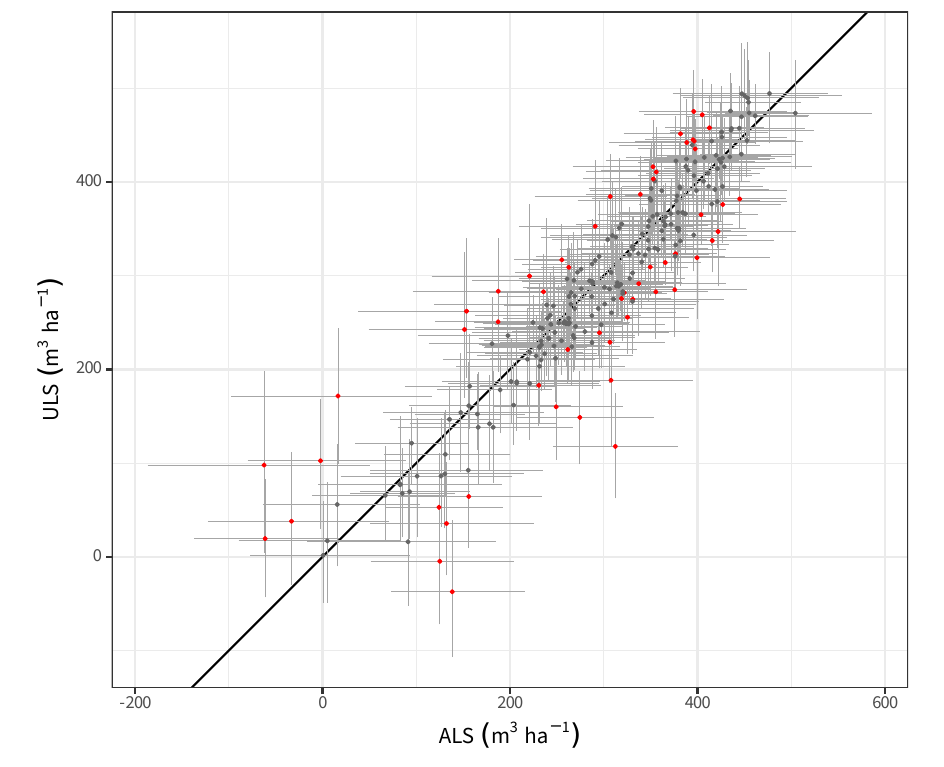}
    \caption{Stand-level posterior mean predictions for ULS- versus ALS-based spatial models at \qty{3}{\meter} resolution. Error bars indicate 95\% credible intervals for each stand. Stands highlighted in red exhibit a 95\% credible interval for the stand-specific difference $\eta_{\mathrm{ULS}} - \eta_{\mathrm{ALS}}$ that excludes zero. The solid line indicates the 1:1 identity line.}
    \label{fig:pred_als_vs_uls}
\end{figure}

A direct stand-by-stand comparison between the two models is shown in Figure~\ref{fig:pred_als_vs_uls}. Predictions align closely with the 1:1 line, indicating that no strong systematic difference exists between the ALS- and ULS-based stand-level posterior means. Nevertheless, for the highlighted subset of stands, the marginal 95\% credible interval for the paired difference $\eta_{\mathrm{ULS}} - \eta_{\mathrm{ALS}}$ excludes zero, indicating substantive local discrepancies in predicted GSV between the two technologies. Slightly negative mean predictions occurred for \num{2} stands in the ULS model and \num{3} stands in the ALS model. These instances restricted to very young, regenerating stands where tree dimensions remained below the \qty{10}{\centi\meter} DBH inventory threshold.

\begin{figure}
    \centering
    \includegraphics[width=1\linewidth]{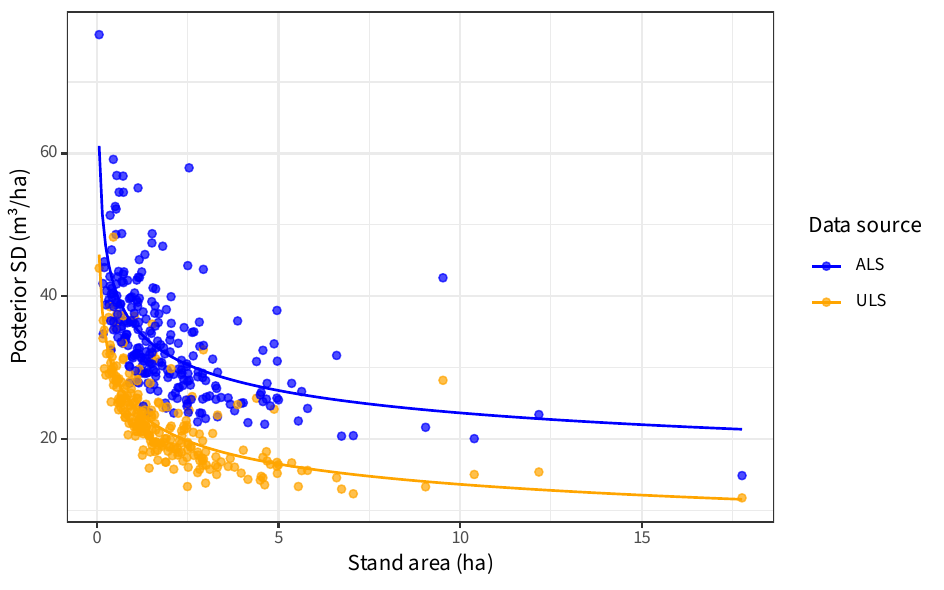}
    \caption{Stand-level posterior SD as a function of stand area for \qty{3}{\meter} full-predictor models using ULS and ALS LiDAR data, with lines showing descriptive fitted power-law decay curves ($SD = \tau + a/\mathrm{Area}^p$) shown per platform.}
    \label{fig:error_size}
\end{figure}

Figure~\ref{fig:error_size} illustrates how posterior SD relates to stand area. Posterior uncertainty tended to decrease with larger stand sizes, though this trend was moderate and appeared primarily as a reduction in maximum uncertainty values. Small stands exhibited both the highest and most variable posterior SDs, whereas larger stands produced lower, more stable uncertainty bounds. This behavior aligns with spatial aggregation effects reducing variance in stand-level means, though the persistent scatter confirms that structural conditions and stand volume also exert strong influences on stand-level precision. 

\begin{figure}
    \centering
    \includegraphics[width=1\linewidth]{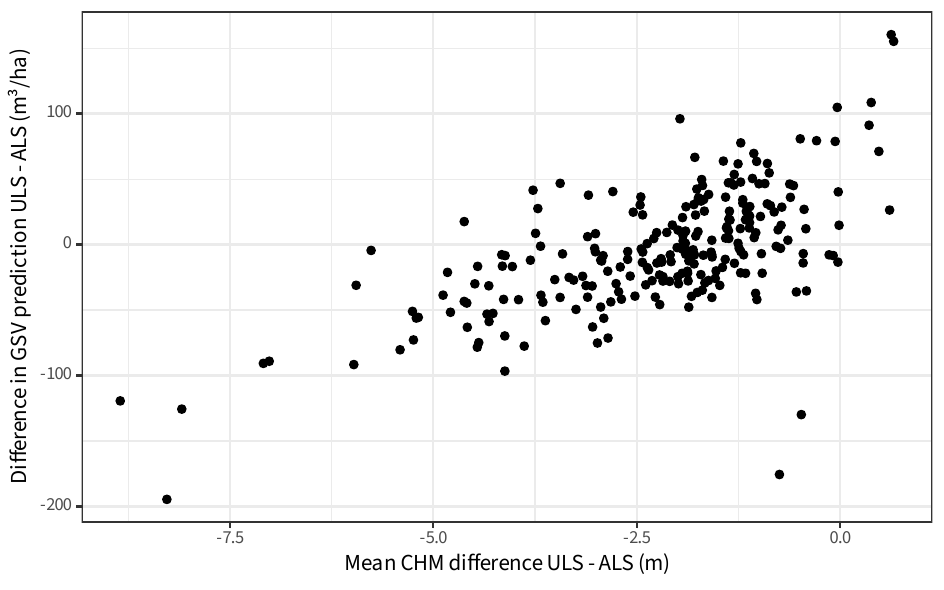}
    \caption{Relationship between stand-wise differences in mean CHM height ($\mathrm{ULS} - \mathrm{ALS}$) and corresponding differences in predicted posterior mean GSV.}
    \label{fig:chm_diff}
\end{figure}

To clarify the drivers behind these discrepancies, stand-wise differences in predicted GSV were evaluated against corresponding differences in mean CHM height ($\mathrm{ULS} - \mathrm{ALS}$). As shown in Figure~\ref{fig:chm_diff}, differences in predicted GSV were positively correlated with CHM height ($r = 0.64$). This relationship demonstrates that acquisition-related canopy height differences partly explain the divergence between ALS- and ULS-based GSV predictions. In particular, stands with markedly negative CHM differences, which typically correspond to thinnings or canopy removals occurring between sensor flights, yielded lower GSV predictions under ULS than under ALS.

\subsection{Population-level inference}
\begin{table}
\centering
\caption{\label{tab:estimators}Population-level estimates for growing stock volume
(\unit{\cubic\meter\per\hectare}). Stratifications are A: by previous stand
inventory dominant leaf type (two strata), B: by CORINE dominant leaf type
(two strata), C: by previous stand inventory age class (four strata).}
\centering
\small
\begin{tabular}[t]{lrrrl}
\toprule
\textbf{Estimator} & \textbf{Mean} & \textbf{SE/SD} & \textbf{RSE\%/RSD\%} & \textbf{95\% CI}\\
\midrule
\addlinespace[0.3em]
\multicolumn{5}{l}{\textbf{Design-based}}\\
\hspace{1em}Horvitz-Thompson & 330.6 & 10.9 & 3.3 & (309.1, 352.1)\\
\hspace{1em}Stratification A & 320.1 & 9.7 & 3.0 & (301.0, 339.2)\\
\hspace{1em}Stratification B & 319.9 & 10.0 & 3.1 & (300.1, 339.7)\\
\hspace{1em}Stratification C & 321.7 & 9.7 & 3.0 & (302.6, 340.8)\\
\hspace{1em}Regression Estimator & 304.7 & 6.3 & 2.1 & (292.3, 317.0)\\
\addlinespace[0.3em]
\multicolumn{5}{l}{\textbf{Model-based}}\\
\hspace{1em}ALS & 314.5 & 5.8 & 1.9 & (303.1, 326.0)\\
\hspace{1em}ULS & 309.6 & 5.7 & 1.8 & (298.0, 320.5)\\
\bottomrule
\end{tabular}
\end{table}
 Population-level estimates for the study area are summarized in Table~\ref{tab:estimators}. Design-based estimators of yielded GSV values ranging from \num{304.7} to \qty{330.6}{\cubic\meter\per\hectare}. The regression estimator produced a lower overall mean than the SYS estimator, representing a reduction of \qty{7.8}{\percent}. This shift may likely reflects minor operational deviations from a perfect probabilistic grid, as omitted inventory plots were concentrated in marshy areas with characteristically low growing stock. Post-stratified estimators aligned closely with the unstratified SYS mean.

None of the three post-stratification schemes achieved a notable reduction in sampling error, which indicates that neither age-class nor canopy-type strata capture major deviations from the global population mean. In contrast, the regression estimator substantially improved design-based precision, reducing the standard error by \qty{42.2}{\percent} compared to the SYS baseline.

The model-based population means were \qty{314.5}{\cubic\meter\per\hectare} for the ALS platform and \qty{309.6}{\cubic\meter\per\hectare} for the ULS platform. Taking the regression estimator as the primary design-based benchmark, both model-based estimates fell well within its \qty{95}{\percent} confidence interval. The Bayesian credible intervals were narrower than the design-based confidence intervals, though this difference should be viewed descriptively given the distinct statistical principles underlying the two paradigms and should be based on population-support prediction rather than aggregation of stand-wise marginal draws.

\section{Discussion}
This study evaluated alternative data acquisition strategies and modeling approaches for model-based forest inventories using a Bayesian change-of-support (COS) regression framework. The findings provide practical insights for forest managers and organizations aiming to implement stand-level estimation of GSV. Our results demonstrate that Bayesian change-of-support models enhance predictive performance by allowing for high resolution prediction grids, particularly when prediction targets such as forest stands are small or irregularly shaped. We also quantified the added value of high-density drone-based LiDAR (ULS) compared with operational airborne laser scanning (ALS). Finally, the analysis indicates that under the forest conditions examined here, non-spatial models provided sufficient predictive power without requiring computationally intensive spatial random fields.

The adoption of high resolution COS models improved GSV prediction precision relative to coarser-support alternatives, yielding substantial reductions in cross-validated prediction errors. This improvement aligns with the premise that conventional raster-based models suffer from spatial mismatch, where field observations and fine-scale auxiliary metrics operate at different spatial scales. COS modeling is therefore especially advantageous when target units are small or exhibit high perimeter-to-area ratios. In such cases, coarse prediction pixels that overlap stand boundaries induce spatial smoothing, which biases stand estimates toward the surrounding landscape mean. Fine-resolution auxiliary and prediction grids, enabled by the COS architecture, effectively mitigate edge mismatch and spatial averaging. However, generating reliable high-resolution predictor rasters requires adequate laser point density. In this study, distributional LiDAR metrics at a \qty{3}{\meter} resolution improved predictions only for ULS-based models, while the lower density of ALS data resulted in missing cell-wise statistics across more than \qty{1}{\percent} of the study area.

Parameter estimates were broadly consistent with ecological expectations: GSV increased with forest height (\textit{chm}) and canopy density (\textit{pct\_above2}) across all model variants. The interpretation of heterogeneity (\textit{cv\_z}) differed between spatial scales. At \qty{3}{\meter}, it had a positive association with GSV, possibly reflecting vertical layering and thus greater local biomass. At \qty{35}{\meter}, however, the association was negative, suggesting that heterogeneity at this scale may instead capture canopy gaps and horizontal structural variation. The effect of \textit{pct\_10\_20}, a proxy for understorey density, was more difficult to interpret, as its credible interval did not exclude zero in most model variants. Parameter estimates should nevertheless be interpreted cautiously because of collinearity among structural LiDAR metrics and nonlinear relationships across spatial scales.

Regarding data acquisition, the combination of elevated point density and a tighter temporal alignment with field measurements gave ULS models a clear advantage over ALS. Across all model structures, ULS data consistently yielded lower cross-validated prediction errors and narrower plot-level prediction intervals. For operational forest inventories, however, this performance gain must be weighted against the additional cost of dedicated drone flights compared to using existing regional ALS coverages. Our findings provide a quantitative basis for evaluating this trade-off in similar temperate forest settings.

Due to the observational design, the effects of lower point density and temporal mismatch could not be completely disentangled. Even after removing plots with the largest acquisition lag, residual differences between ULS and ALS predictions remained correlated with indicators of forest structural change, such as CHM height differences. This pattern suggests that part of the elevated error in ALS models stems from unmeasured management interventions, such as thinnings or final harvests, occurring between the remote sensing flight and yield inventory. Concurrently, differences in canopy penetration and sampling geometry between ALS and ULS influenced predictor quality. Compared to ULS, ALS models exhibited altered posterior parameter estimates, wider credible intervals, and reduced sensitivity to distributional metrics, reflecting structural differences in data quality beyond simple acquisition delay. Notwithstanding the methodological limitations, comparing older ALS data with more recent ULS data reflects a common operational reality: ALS acquisitions are often commissioned by external providers and may not align with forest planning or inventory intervals.

Under the conditions investigated, spatial models only provided a small improvement in comparison with non-spatial models once LiDAR-derived covariates were included, especially when using ULS data. Auxiliary metrics appear to have absorbed much of the spatially structured variation relevant for prediction, so the residual spatial process contributed little to out-of-sample predictive performance even though the fitted spatial variance component was not zero. The slightly larger performance gain in ALS models likely reflects localized, spatially auto-correlated errors induced by management activity between flight and inventory dates. Clustered harvesting operations, for example, create spatial patches where remote sensing predictors could systematically over-predict current volume. Furthermore, the \qty{183}{\meter} sample grid spacing restricts the capacity to detect residual spatial autocorrelation at sub-grid scales, such as micro-topographic variations or intra-stand homogeneity. The limited additional benefit of spatial modeling aligns with the findings of \citet{Nothdurft_2025}, who likewise found that simpler non-spatial models performed comparably to more complex spatial variants when the predictors already explained a large proportion of the variance. Although non-spatial models were adequate for the present study area, stronger residual spatial structure may emerge in more complex terrain, contiguous natural forests, or under sparser sampling designs.

Although personal laser scanning (PLS) is not strictly required for model-based inference, it offered distinct operational benefits over more traditional fixed-radius plot surveys. Following field scanning, the automated processing pipeline enabled rapid plot establishment, exact tree georeferencing, and a single-operator workflow. The generated point clouds also provided direct tree height measurements, reducing reliance on height-diameter regression models, although \qty{8.6}{\percent} of tree heights still required model imputation due to local occlusion or incomplete canopy penetration. Furthermore, while real-time kinematic (RTK) positioning achieved multi-meter accuracy beneath closed canopies, co-registered local PLS canopy height models with airborne rasters enabled fine-scale alignment, resulting in positional accuracy not achievable with GNSS positioning alone. This improved spatial co-registration reduces positional error in empirical models \citep{frazer_simulated_2011} and establishes precise tree-level coordinates necessary for repeat forest-inventory tracking.

\section{Conclusion}
This study demonstrates the effectiveness of a Bayesian change-of-support (COS) modeling framework for stand-level forest inventory in a structurally complex, species-rich forest using multi-platform laser scanning data. By explicitly accounting for spatial support mismatch between field sample plots and high-resolution predictor grids, the COS approach significantly improves the accuracy of growing stock volume (GSV) predictions. Operating at a \qty{3}{\meter} resolution effectively mitigates spatial averaging errors, which is particularly beneficial for small, irregularly shaped forest management units.

From a data acquisition perspective, high-density drone-based LiDAR (ULS) provided clear advantages over operational airborne laser scanning (ALS). The superior point density of ULS enabled the effective use of fine-scale distributional metrics and, combined with closer temporal proximity to field inventories, yielded lower cross-validated prediction errors and narrower plot-level prediction intervals. However, for operational applications, managers must weigh these performance gains against the additional costs of dedicated UAV surveys compared to existing regional ALS coverages.

Crucially, our findings indicate that once rich structural LiDAR metrics from ULS are incorporated, non-spatial model variants achieve predictive performance comparable to computationally demanding spatial models. Furthermore, model-based population estimates aligned closely with design-based regression benchmarks and produced narrower model-based intervals, although these intervals should be interpreted within the Bayesian model rather than as direct design-based uncertainty estimates. Overall, the integration of personal laser scanning for plot surveys, high-density aerial LiDAR, and Bayesian change-of-support modeling offers a robust and scalable solution for modern, precision-oriented forest inventory and forest management.

\section*{Author contributions}
\textbf{Arthur Spitta:} Conceptualization, Methodology, Investigation, Data curation, Formal Analysis, Software, Visualization, Writing --- original draft, Writing --- review \& editing. \textbf{Andrew O. Finley:} Methodology, Software, Validation, Writing --- original draft, Writing --- review \& editing. \textbf{Christoph Gollob:} Writing --- review \& editing. \textbf{Ralf 
Kraßnitzer:} Resources, Writing --- review \& editing. \textbf{Tim Ritter:} Resources; Data curation; Writing --- review \& editing. \textbf{Andreas Tockner:} Data curation, Software, Writing --- review \& editing. \textbf{Arne Nothdurft:} Conceptualization, Funding Acquisition, Methodology, Writing --- review \& editing, Project administration, Supervision.

\section*{Declaration of generative AI use}
During the preparation of this manuscript, the authors used used OpenAI's GPT-5 and GPT-5.6 for language editing, structural refinement, and phrasing support. The authors thoroughly reviewed, revised, and verified all outputs and take full responsibility for the final content of the manuscript.

\section*{Declaration of competing interests}
The authors declare there are no competing interests.

\section*{Funding}
This research was supported by the State Forestry Agency of Mecklenburg–West Pomerania (Landesforstanstalt Mecklenburg-Vorpommern).

\section*{Data availability}
The field inventory data used in this study cannot be made publicly available due to legal and privacy restrictions. The analytical code and R scripts are available from the corresponding author upon reasonable request.

\section*{Acknowledgments}
We would like to express our sincere gratitude to Kai Jütte for initiating this research initiative and for his continuous guidance and support throughout the project.

We also thank OGF GmbH for acquiring and providing the UAV-borne laser scanning (ULS) data used in this study on behalf of the State Forestry Agency of Mecklenburg–West Pomerania.
\bibliography{references_abbrev}

\clearpage
\begin{appendices}

\renewcommand{\thetable}{A\arabic{table}}
\renewcommand{\thefigure}{A\arabic{figure}}

\setcounter{table}{0}
\setcounter{figure}{0}

\section{Detailed posterior model summaries}
\label{sec:appendix_models}

\clearpage
\newgeometry{margin=1cm}
\begin{landscape}
\thispagestyle{empty}
\begin{table}
\centering
\caption{Posterior parameter estimates for the 3 m full-predictor models. Values are posterior medians with 95\% credible intervals in parentheses.}
\label{tab:pp_full_3_app}
\centering
\footnotesize
\begin{tabular}[t]{lllll}
\toprule
\multicolumn{1}{c}{\textbf{}} & \multicolumn{2}{c}{\textbf{ULS}} & \multicolumn{2}{c}{\textbf{ALS}} \\
\cmidrule(l{3pt}r{3pt}){2-3} \cmidrule(l{3pt}r{3pt}){4-5}
\multicolumn{1}{c}{Parameter} & \multicolumn{1}{c}{Spatial} & \multicolumn{1}{c}{Nonspatial} & \multicolumn{1}{c}{Spatial} & \multicolumn{1}{c}{Nonspatial}\\
\midrule
$\beta_0$ & \makebox[4.2em][r]{\num{319.2}}\;(\num{307.7},\,\num{330.5}) & \makebox[4.2em][r]{\num{318.4}}\;(\num{306.4},\,\num{329.3}) & \makebox[4.2em][r]{\num{315.4}}\;(\num{301.2},\,\num{332.7}) & \makebox[4.2em][r]{\num{314.0}}\;(\num{300.4},\,\num{327.0})\\
$\beta_{\mathrm{chm}}$ & \makebox[4.2em][r]{\num{89.0}}\;(\num{70.9},\,\num{105.1}) & \makebox[4.2em][r]{\num{88.8}}\;(\num{72.3},\,\num{105.1}) & \makebox[4.2em][r]{\num{90.4}}\;(\num{72.2},\,\num{106.9}) & \makebox[4.2em][r]{\num{90.0}}\;(\num{71.2},\,\num{108.1})\\
$\beta_{\mathrm{pct\,above2}}$ & \makebox[4.2em][r]{\num{37.7}}\;(\num{25.4},\,\num{50.6}) & \makebox[4.2em][r]{\num{37.5}}\;(\num{24.4},\,\num{51.0}) & \makebox[4.2em][r]{\num{20.2}}\;(\num{2.0},\,\num{37.9}) & \makebox[4.2em][r]{\num{19.8}}\;(\num{0.6},\,\num{38.8})\\
$\beta_{\mathrm{pct\,10\,20}}$ & \makebox[4.2em][r]{\num{-9.6}}\;(\num{-22.4},\,\num{4.0}) & \makebox[4.2em][r]{\num{-9.8}}\;(\num{-24.4},\,\num{4.3}) & \makebox[4.2em][r]{\num{5.8}}\;(\num{-9.3},\,\num{20.8}) & \makebox[4.2em][r]{\num{6.3}}\;(\num{-9.2},\,\num{21.5})\\
$\beta_{\mathrm{cv\,z}}$ & \makebox[4.2em][r]{\num{18.7}}\;(\num{4.6},\,\num{32.2}) & \makebox[4.2em][r]{\num{19.1}}\;(\num{5.3},\,\num{32.2}) & \makebox[4.2em][r]{\num{9.8}}\;(\num{-4.8},\,\num{23.3}) & \makebox[4.2em][r]{\num{11.2}}\;(\num{-3.1},\,\num{24.5})\\
$\tau_B^2$ & \makebox[4.2em][r]{\num{3425.1}}\;(\num{885.4},\,\num{5095.7}) & \makebox[4.2em][r]{\num{4630.7}}\;(\num{3687.4},\,\num{5815.3}) & \makebox[4.2em][r]{\num{3881.0}}\;(\num{917.4},\,\num{6882.9}) & \makebox[4.2em][r]{\num{6287.3}}\;(\num{5047.1},\,\num{7910.0})\\
$\tau^2$ & \makebox[4.2em][r]{\num{458082.5}}\;(\num{118419.7},\,\num{681504.3}) & \makebox[4.2em][r]{\num{619316.3}}\;(\num{493162.3},\,\num{777744.4}) & \makebox[4.2em][r]{\num{514061.2}}\;(\num{121518.5},\,\num{911676.0}) & \makebox[4.2em][r]{\num{832784.9}}\;(\num{668521.8},\,\num{1047721.5})\\
$\sigma^2$ & \makebox[4.2em][r]{\num{2217.8}}\;(\num{578.6},\,\num{10247.4}) & \multicolumn{1}{c}{\textemdash{}} & \makebox[4.2em][r]{\num{2993.1}}\;(\num{712.6},\,\num{10122.7}) & \multicolumn{1}{c}{\textemdash{}}\\
Eff. range (m) & \makebox[4.2em][r]{\num{59.6}}\;(\num{30.7},\,\num{465.0}) & \multicolumn{1}{c}{\textemdash{}} & \makebox[4.2em][r]{\num{279.3}}\;(\num{32.2},\,\num{940.6}) & \multicolumn{1}{c}{\textemdash{}}\\
\midrule
$R^2$ & \makebox[4.2em][r]{\num{0.78}}\;(\num{0.68},\,\num{0.92}) & \makebox[4.2em][r]{\num{0.70}}\;(\num{0.62},\,\num{0.76}) & \makebox[4.2em][r]{\num{0.74}}\;(\num{0.59},\,\num{0.92}) & \makebox[4.2em][r]{\num{0.59}}\;(\num{0.50},\,\num{0.66})\\
\bottomrule
\end{tabular}
\end{table}

 \begin{table}
\centering
\caption{Posterior parameter estimates for the 35 m full-predictor models. Values are posterior medians with 95\% credible intervals in parentheses.}
\label{tab:pp_full_35_app}
\centering
\footnotesize
\begin{tabular}[t]{lllll}
\toprule
\multicolumn{1}{c}{\textbf{}} & \multicolumn{2}{c}{\textbf{ULS}} & \multicolumn{2}{c}{\textbf{ALS}} \\
\cmidrule(l{3pt}r{3pt}){2-3} \cmidrule(l{3pt}r{3pt}){4-5}
\multicolumn{1}{c}{Parameter} & \multicolumn{1}{c}{Spatial} & \multicolumn{1}{c}{Nonspatial} & \multicolumn{1}{c}{Spatial} & \multicolumn{1}{c}{Nonspatial}\\
\midrule
$\beta_0$ & \makebox[4.2em][r]{\num{322.8}}\;(\num{309.0},\,\num{338.8}) & \makebox[4.2em][r]{\num{320.3}}\;(\num{305.5},\,\num{334.0}) & \makebox[4.2em][r]{\num{322.5}}\;(\num{306.2},\,\num{343.9}) & \makebox[4.2em][r]{\num{318.9}}\;(\num{302.1},\,\num{333.7})\\
$\beta_{\mathrm{chm}}$ & \makebox[4.2em][r]{\num{49.6}}\;(\num{21.7},\,\num{73.4}) & \makebox[4.2em][r]{\num{49.0}}\;(\num{23.5},\,\num{74.1}) & \makebox[4.2em][r]{\num{41.7}}\;(\num{15.4},\,\num{66.4}) & \makebox[4.2em][r]{\num{41.8}}\;(\num{14.0},\,\num{65.8})\\
$\beta_{\mathrm{pct\,above2}}$ & \makebox[4.2em][r]{\num{39.3}}\;(\num{23.8},\,\num{55.7}) & \makebox[4.2em][r]{\num{39.4}}\;(\num{23.4},\,\num{55.8}) & \makebox[4.2em][r]{\num{28.6}}\;(\num{9.5},\,\num{47.6}) & \makebox[4.2em][r]{\num{24.5}}\;(\num{5.1},\,\num{43.0})\\
$\beta_{\mathrm{pct\,10\,20}}$ & \makebox[4.2em][r]{\num{-24.4}}\;(\num{-42.9},\,\num{-5.5}) & \makebox[4.2em][r]{\num{-24.9}}\;(\num{-45.4},\,\num{-6.4}) & \makebox[4.2em][r]{\num{-13.0}}\;(\num{-32.3},\,\num{7.2}) & \makebox[4.2em][r]{\num{-10.4}}\;(\num{-30.9},\,\num{9.5})\\
$\beta_{\mathrm{cv\,z}}$ & \makebox[4.2em][r]{\num{-8.4}}\;(\num{-29.2},\,\num{12.1}) & \makebox[4.2em][r]{\num{-7.4}}\;(\num{-28.5},\,\num{12.9}) & \makebox[4.2em][r]{\num{-26.0}}\;(\num{-47.2},\,\num{-6.2}) & \makebox[4.2em][r]{\num{-25.9}}\;(\num{-46.4},\,\num{-6.9})\\
$\tau_B^2$ & \makebox[4.2em][r]{\num{4504.6}}\;(\num{837.4},\,\num{7985.8}) & \makebox[4.2em][r]{\num{7441.4}}\;(\num{6005.2},\,\num{9379.4}) & \makebox[4.2em][r]{\num{4353.9}}\;(\num{737.5},\,\num{9809.1}) & \makebox[4.2em][r]{\num{9379.8}}\;(\num{7470.6},\,\num{11818.5})\\
$\tau^2$ & \makebox[4.2em][r]{\num{9771.2}}\;(\num{1816.5},\,\num{17322.5}) & \makebox[4.2em][r]{\num{16141.6}}\;(\num{13026.3},\,\num{20345.6}) & \makebox[4.2em][r]{\num{9444.2}}\;(\num{1599.9},\,\num{21277.5}) & \makebox[4.2em][r]{\num{20346.4}}\;(\num{16205.0},\,\num{25636.3})\\
$\sigma^2$ & \makebox[4.2em][r]{\num{3780.4}}\;(\num{730.3},\,\num{12348.4}) & \multicolumn{1}{c}{\textemdash{}} & \makebox[4.2em][r]{\num{5782.9}}\;(\num{812.9},\,\num{14838.4}) & \multicolumn{1}{c}{\textemdash{}}\\
Eff. range (m)& \makebox[4.2em][r]{\num{103.6}}\;(\num{31.1},\,\num{783.1}) & \multicolumn{1}{c}{\textemdash{}} & \makebox[4.2em][r]{\num{309.4}}\;(\num{33.0},\,\num{913.4}) & \multicolumn{1}{c}{\textemdash{}}\\
\midrule
$R^2$ & \makebox[4.2em][r]{\num{0.71}}\;(\num{0.55},\,\num{0.93}) & \makebox[4.2em][r]{\num{0.53}}\;(\num{0.43},\,\num{0.61}) & \makebox[4.2em][r]{\num{0.72}}\;(\num{0.46},\,\num{0.94}) & \makebox[4.2em][r]{\num{0.40}}\;(\num{0.29},\,\num{0.50})\\
\bottomrule
\end{tabular}
\end{table}

 \end{landscape}
\restoregeometry

\clearpage
\newgeometry{margin=1cm}
\begin{landscape}
\thispagestyle{empty}
\begin{table}
\centering
\caption{Posterior parameter estimates for the 3 m intercept-only and CHM-only models. Values are posterior medians with 95\% credible intervals in parentheses.}
\label{tab:pp_red_3_app}
\centering
\footnotesize
\begin{tabular}[t]{llllll}
\toprule
\multicolumn{1}{c}{\textbf{}} & \multicolumn{1}{c}{\textbf{}} & \multicolumn{2}{c}{\textbf{ULS}} & \multicolumn{2}{c}{\textbf{ALS}} \\
\cmidrule(l{3pt}r{3pt}){3-4} \cmidrule(l{3pt}r{3pt}){5-6}
\multicolumn{1}{c}{Parameter} & \multicolumn{1}{c}{Intercept-only spatial} & \multicolumn{1}{c}{Spatial} & \multicolumn{1}{c}{Nonspatial} & \multicolumn{1}{c}{Spatial} & \multicolumn{1}{c}{Nonspatial}\\
\midrule
$\beta_0$ & \makebox[4.2em][r]{\num{324.2}}\;(\num{303.5},\,\num{346.2}) & \makebox[4.2em][r]{\num{323.8}}\;(\num{311.6},\,\num{336.7}) & \makebox[4.2em][r]{\num{323.3}}\;(\num{311.1},\,\num{335.4}) & \makebox[4.2em][r]{\num{326.0}}\;(\num{310.0},\,\num{344.8}) & \makebox[4.2em][r]{\num{322.7}}\;(\num{309.3},\,\num{335.8})\\
$\beta_{\mathrm{chm}}$ & \multicolumn{1}{c}{\textemdash{}} & \makebox[4.2em][r]{\num{96.5}}\;(\num{84.3},\,\num{109.1}) & \makebox[4.2em][r]{\num{96.2}}\;(\num{83.3},\,\num{109.1}) & \makebox[4.2em][r]{\num{95.9}}\;(\num{82.9},\,\num{107.9}) & \makebox[4.2em][r]{\num{93.3}}\;(\num{80.4},\,\num{106.2})\\
$\tau_B^2$ & \makebox[4.2em][r]{\num{9372.4}}\;(\num{1601.6},\,\num{16298.5}) & \makebox[4.2em][r]{\num{4262.2}}\;(\num{1411.8},\,\num{6185.0}) & \makebox[4.2em][r]{\num{5684.3}}\;(\num{4544.0},\,\num{7259.9}) & \makebox[4.2em][r]{\num{3295.5}}\;(\num{967.1},\,\num{6275.1}) & \makebox[4.2em][r]{\num{6181.4}}\;(\num{4995.7},\,\num{7834.9})\\
$\tau^2$ & \makebox[4.2em][r]{\num{1270718.9}}\;(\num{217145.7},\,\num{2209758.9}) & \makebox[4.2em][r]{\num{577871.5}}\;(\num{191410.3},\,\num{838570.1}) & \makebox[4.2em][r]{\num{770681.7}}\;(\num{616082.1},\,\num{984299.8}) & \makebox[4.2em][r]{\num{446799.4}}\;(\num{131114.8},\,\num{850776.0}) & \makebox[4.2em][r]{\num{838080.6}}\;(\num{677318.7},\,\num{1062258.4})\\
$\sigma^2$ & \makebox[4.2em][r]{\num{8639.0}}\;(\num{1780.2},\,\num{32261.7}) & \makebox[4.2em][r]{\num{2487.7}}\;(\num{715.1},\,\num{10524.3}) & \multicolumn{1}{c}{\textemdash{}} & \makebox[4.2em][r]{\num{3306.7}}\;(\num{1077.3},\,\num{8003.4}) & \multicolumn{1}{c}{\textemdash{}}\\
Eff. range (m)& \makebox[4.2em][r]{\num{114.6}}\;(\num{31.9},\,\num{673.3}) & \makebox[4.2em][r]{\num{68.7}}\;(\num{31.0},\,\num{685.3}) & \multicolumn{1}{c}{\textemdash{}} & \makebox[4.2em][r]{\num{508.5}}\;(\num{33.3},\,\num{965.6}) & \multicolumn{1}{c}{\textemdash{}}\\
\midrule
$R^2$ & \makebox[4.2em][r]{\num{0.39}}\;(\num{0.08},\,\num{0.81}) & \makebox[4.2em][r]{\num{0.72}}\;(\num{0.62},\,\num{0.89}) & \makebox[4.2em][r]{\num{0.62}}\;(\num{0.53},\,\num{0.69}) & \makebox[4.2em][r]{\num{0.78}}\;(\num{0.64},\,\num{0.93}) & \makebox[4.2em][r]{\num{0.58}}\;(\num{0.49},\,\num{0.67})\\
\bottomrule
\end{tabular}
\end{table}
 \begin{table}
\centering
\caption{Posterior parameter estimates for the 35 m intercept-only and CHM-only models. Values are posterior medians with 95\% credible intervals in parentheses.}
\label{tab:pp_red_35_app}
\centering
\footnotesize
\begin{tabular}[t]{llllll}
\toprule
\multicolumn{1}{c}{\textbf{}} & \multicolumn{1}{c}{\textbf{}} & \multicolumn{2}{c}{\textbf{ULS}} & \multicolumn{2}{c}{\textbf{ALS}} \\
\cmidrule(l{3pt}r{3pt}){3-4} \cmidrule(l{3pt}r{3pt}){5-6}
\multicolumn{1}{c}{Parameter} & \multicolumn{1}{c}{Intercept-only spatial} & \multicolumn{1}{c}{Spatial} & \multicolumn{1}{c}{Nonspatial} & \multicolumn{1}{c}{Spatial} & \multicolumn{1}{c}{Nonspatial}\\
\midrule
$\beta_0$ & \makebox[4.2em][r]{\num{320.9}}\;(\num{298.8},\,\num{342.2}) & \makebox[4.2em][r]{\num{321.4}}\;(\num{306.5},\,\num{338.6}) & \makebox[4.2em][r]{\num{319.4}}\;(\num{303.9},\,\num{335.0}) & \makebox[4.2em][r]{\num{321.6}}\;(\num{304.5},\,\num{339.8}) & \makebox[4.2em][r]{\num{318.6}}\;(\num{302.7},\,\num{334.7})\\
$\beta_{\mathrm{chm}}$ & \multicolumn{1}{c}{\textemdash{}} & \makebox[4.2em][r]{\num{82.4}}\;(\num{67.7},\,\num{96.7}) & \makebox[4.2em][r]{\num{81.5}}\;(\num{66.6},\,\num{96.5}) & \makebox[4.2em][r]{\num{75.2}}\;(\num{59.0},\,\num{91.5}) & \makebox[4.2em][r]{\num{74.1}}\;(\num{57.0},\,\num{90.5})\\
$\tau_B^2$ & \makebox[4.2em][r]{\num{6630.1}}\;(\num{1689.0},\,\num{15824.4}) & \makebox[4.2em][r]{\num{5282.5}}\;(\num{1262.0},\,\num{9171.5}) & \makebox[4.2em][r]{\num{8680.5}}\;(\num{6915.3},\,\num{11091.5}) & \makebox[4.2em][r]{\num{5522.6}}\;(\num{1269.7},\,\num{10415.6}) & \makebox[4.2em][r]{\num{9992.3}}\;(\num{8103.7},\,\num{12689.8})\\
$\tau^2$ & \makebox[4.2em][r]{\num{14381.8}}\;(\num{3663.7},\,\num{34325.9}) & \makebox[4.2em][r]{\num{11458.7}}\;(\num{2737.5},\,\num{19894.6}) & \makebox[4.2em][r]{\num{18829.4}}\;(\num{15000.4},\,\num{24059.4}) & \makebox[4.2em][r]{\num{11979.5}}\;(\num{2754.3},\,\num{22593.2}) & \makebox[4.2em][r]{\num{21675.0}}\;(\num{17578.3},\,\num{27526.3})\\
$\sigma^2$ & \makebox[4.2em][r]{\num{11133.9}}\;(\num{2066.3},\,\num{23991.5}) & \makebox[4.2em][r]{\num{4227.7}}\;(\num{931.0},\,\num{13182.8}) & \multicolumn{1}{c}{\textemdash{}} & \makebox[4.2em][r]{\num{5457.9}}\;(\num{1060.1},\,\num{15941.6}) & \multicolumn{1}{c}{\textemdash{}}\\
Eff. range (m) & \makebox[4.2em][r]{\num{197.2}}\;(\num{32.2},\,\num{747.2}) & \makebox[4.2em][r]{\num{143.3}}\;(\num{31.4},\,\num{908.9}) & \multicolumn{1}{c}{\textemdash{}} & \makebox[4.2em][r]{\num{180.3}}\;(\num{32.0},\,\num{907.8}) & \multicolumn{1}{c}{\textemdash{}}\\
\midrule
$R^2$ & \makebox[4.2em][r]{\num{0.54}}\;(\num{0.14},\,\num{0.85}) & \makebox[4.2em][r]{\num{0.66}}\;(\num{0.48},\,\num{0.89}) & \makebox[4.2em][r]{\num{0.43}}\;(\num{0.33},\,\num{0.53}) & \makebox[4.2em][r]{\num{0.64}}\;(\num{0.38},\,\num{0.89}) & \makebox[4.2em][r]{\num{0.35}}\;(\num{0.24},\,\num{0.46})\\
\bottomrule
\end{tabular}
\end{table} \end{landscape}
\restoregeometry

\end{appendices}

\end{document}